\documentclass{aa}

\usepackage{epsfig,graphicx,lscape}
\usepackage{float,longtable,comment}
\usepackage{rotating}
\usepackage{natbib}
\newcommand{\kms}{{{\,km\,s}$^{-1}$\,}}
\newcommand{\teff}{{$T_\mathrm{eff}$\,}}
\newcommand{\logg}{{log~$g$\,}}
\newcommand{\vsini}{$v\sin i\,$}
\newcommand{\Msun}{\,$\mathrm{M}_\odot$}

\begin{document}
\title{The VLT-FLAMES survey of massive stars: Nitrogen abundances for Be-type stars in the Magellanic Clouds\thanks{Based on observations at the European Southern
Observatory in programmes 171.0237 and 073.0234}}

\author{P.R. Dunstall\inst{1}, I. Brott\inst{2}, P.L. Dufton\inst{1},  D.J. Lennon\inst{4}, C.J. Evans\inst{3}, S.J. Smartt\inst{1}, I. Hunter\inst{1}
}

\institute{Department of Physics \& Astronomy, The Queen's University 
of Belfast, BT7 1NN, Northern Ireland, UK
	\and University of Vienna, Department of Astronomy, T\"{u}rkenschanzstr. 17, A-1180 Vienna, Austria
	\and UK Astronomy Technology Centre, Royal Observatory Edinburgh, Blackford Hill, Edinburgh, EH9 3HJ, UK
	\and ESA, Space Telescope Science Institute, 3700 San Martin Drive, Baltimore, MD 21218, USA
}
       

\date{Received; accepted }

\abstract{}
{We compare the predictions of evolutionary models for early-type stars with atmospheric parameters, projected rotational velocities and nitrogen abundances estimated for a sample of Be-type stars. Our targets are located in 4 fields centred on the Large Magellanic Cloud cluster: NGC\,2004 and the N\,11 region as well as the Small Magellanic Cloud clusters: NGC\,330 and NGC\,346.  }
{Atmospheric parameters and photospheric abundances  have been determined using  the non-LTE atmosphere code {\sc tlusty}.  Effective temperature estimates were deduced using three different methodologies depending on the  spectral features observed; in general they were found to yield consistent estimates. Gravities were deduced from Balmer line profiles and microturbulences from the \ion{Si}{iii} spectrum.  Additionally the contributions of continuum emission from circumstellar discs were estimated. Given its importance in constraining stellar evolutionary models, nitrogen abundances (or upper limits) were deduced for all the stars analysed.}
{Our nitrogen abundances are inconsistent with those predicted for targets spending most of their main sequence life rotating near to the critical velocity. This is consistent with the results we obtain from modelling the inferred rotational velocity distribution of our sample and of other investigators. We consider a number of possibilities to explain the nitrogen abundances and rotational velocities of our Be-type sample.
}{}


\authorrunning{P.R.Dunstall et al}
\titlerunning{Be-type stars in the Magellanic Clouds}

\maketitle
%
\section{Introduction}                                         \label{s_intro}

The classical Be-type star is a B-type star that shows or has previously shown prominent emission features in its Balmer line spectrum, indicating the presence of a geometrically flattened, circumstellar disc \citep{qui94, qui97}.  Although similar phenomena can be observed in some late O- and early A-type stars, it is more prevalent in B spectral types (\citeauthor{con74} \citeyear{con74} - O-type stars; \citeauthor{her60} \citeyear{her60} - A and B-type emission stars). \citet{jas83} and \citet{zor97} have estimated that the frequency of Galactic stars that show the Be phenomenon is approximately 17\%, with the highest fraction around spectral types B1e -- B2e.  Emission is also often seen in the helium and iron spectra, with silicon and magnesium emission lines being seen in some stars \citep{por03}.  Be-type stars can show a variety of spectral characteristics and some can lose their emission line spectrum and transition to a normal B spectral type \citep{cla03}.  

Although the mechanism of the Be phenomena is still unclear, such stars are believed to rotate with velocities $>40\%$ \citep{cra05} up to $\sim 90\%$ of the critical limit, where the centrifugal force balances that of gravity \citep{tow04}.  The Magellanic Clouds are ideal stellar laboratories to investigate the effects of rapid rotation.  These lower metallicity regimes generally produce stars with weaker stellar winds and this reduces the loss of angular momentum, leading to a larger fraction of rapidly rotating stars.  Additionally at lower metallicity, stars of a given mass are more compact again resulting in larger rotational velocities. Therefore, as discussed by \citet{mae99} and \citet{mae00}, lower metallicity regimes should support a larger fraction of Be-type objects.  \citet{wis06}  found an increase in the proportion of Be to B-type stars with decreasing metallicity, whilst \citet{mar06, mar07} observed a large sample of B and Be-type populations in the Magellanic Clouds  and compared them with a Galactic sample, concluding: ``the lower the metallicity is, the higher the rotational velocities are". This conclusion was supported by \citet{hun08a}, but \citet{kel99} observed no difference in their fraction of Be-type stars for different metallicity regimes as part of a photometric based study. 

The evolutionary status of Be-type stars has been discussed by \citet{zor05}.  By determining the stellar mass and age of 97 Be-type objects they reached two important conclusions:  Be-type stars are not constrained to a single stage of the main-sequence but are present over the whole of this evolutionary phase, and more massive stars (M $\ge$ 12\Msun) present the Be phenomena typically earlier in their evolution than less massive stars.  They conclude that this arises from the high mass-loss rates of more massive objects that can reduce their speed of rotation, and also the time scales involved for transferring angular momentum from the core to the surface, which is longer for lower mass stars.

The sample of Be-type stars considered here have been briefly discussed as part of a larger B-type star survey by \citet{hun08a, hun09a}, who undertook a qualitative comparison with the results of \citet{mar06, mar07}. In general good agreement was found e.g. more Be-type stars are present in the SMC than LMC and their projected rotational velocities are higher than B-type stars.  This paper  develops and extends the work of Hunter et al. by estimating the atmospheric parameters and chemical compositions for this sample. 

Nitrogen can be utilised to investigate the degree of mixing of nucleosynthetically processed material from the stellar core into the atmosphere.  Such material shows large enhancements in its nitrogen content, together with smaller changes in the amounts of carbon, oxygen and sodium, other lighter elements such as boron, lithium and beryllium can also be affected \citep{heg00, prz10, bro11a}.  Hence nitrogen is ideal for testing stellar evolutionary models that include the effects of rotational mixing  and here we estimate nitrogen abundances within a sample of 61 Be-type stars from observation fields centred on four open clusters, two in the LMC and two in the SMC.  In Sect. \ref{s_obs}, we outline the observational data collected as part of the VLT-FLAMES survey of massive stars \citep{eva05}.  Sect. \ref{s_analysis} discusses the procedures used to obtain atmospheric parameters and abundance estimates, together with a description of the methodology used to correct our data for contamination due to a circumstellar disc.  Our results are presented and discussed in Sect. \ref{s_discussion} and \ref{s_evo} 
 
\section{Observations}                                        \label{s_obs}

The spectroscopic data were obtained during a European Southern Observatory (ESO) large programme \citep{eva05} using the Fibre Large Array Multi-Element Spectrograph (FLAMES) on the Very Large Telescope.   Observations were obtained for seven clusters within three different metallicity regimes (Galactic, LMC and SMC).  A summary of the spectroscopic configurations is presented in Table \ref{t_obs}, with full details of target selection, data reduction and observational details being given in \citet{eva05, eva06}.

\begin{table*}
\caption{Wavelength coverage, mean FWHM of the arc lines and effective resolving power, R for each Giraffe central wavelength setting, $\lambda_\mathrm{c}$ towards NGC\,346, taken from \citet{eva06}.}
\label{t_obs}
\begin{center}
\begin{tabular}{lccccc}
\hline\hline
 Setting & $\lambda_\mathrm{c}$ & Wavelength Coverage & \multicolumn{2}{c}{FWHM} & R \\
	    &   (\AA)         &          (\AA)  & (\AA)      &  (pixels) & \\
\hline
HR02 & 3958 & 3954 -- 4051 & 0.18 & 3.7 & 22\,000 \\
HR03 & 4124 & 4032 -- 4203 & 0.15 & 3.4 & 27\,500 \\
HR04 & 4297 & 4187 -- 4394 & 0.19 & 3.5 & 22\,600 \\
HR05 & 4471 & 4340 -- 4587 & 0.23 & 3.7 & 19\,450 \\
HR06 & 4656 & 4537 -- 4760 & 0.20 & 3.5 & 23\,300 \\
HR14 & 6515 & 6308 -- 6701 & 0.39 & 3.9 & 16\,700 \\
\hline
\end{tabular}
\end{center}
\end{table*}

\subsection{Selection criteria}			\label{s_crit}

The focus of previous papers has been O-type \citep{mok05, mok06, mok07}, narrow lined B-type stars \citep{tru07, hun07} or fast rotating B-type stars \citep[see][and references therein]{hun09a}.  This paper will provide an analysis of the Be-type stars that were previously omitted from these investigations.  For the Galactic clusters (NGC\,3293, NGC\,4755 and NGC\,6611), no such objects were observed and therefore all our targets lie within the Magellanic Clouds.  The sample is summarised in Table \ref{t_sample}, the spectral type and projected rotational velocity (\vsini) being taken from \citet{hun08a} unless otherwise stated.
  
\begin{table*}
\caption{List of Be-type stars observed in the Magellanic Cloud clusters \object{NGC 346}, \object{NGC 330}, \object{NGC 2004} and the N\,11 region.}
\label{t_sample}
\begin{center}
\begin{tabular}{lcccc|lcccc}
\hline\hline
Star & Spectral & S/N & \vsini & Note & Star & Spectral & S/N & \vsini & Note \\
 & Type & & (\kms) & & & Type & & (\kms) & \\
\hline
NGC\,346-004 & Be(B1:)		& 307 & 266 & ... &		NGC\,330-065 & B1-3e		& 39 & 284 & 1 \\
NGC\,346-008 & B1e		& 187 & 299 & ... &             	NGC\,330-068 & B1.5e 		& 30 & 23  & 1 \\
NGC\,346-009\tablefootmark{*} & B0e	& 114 & 181 & ... &		NGC\,330-069 & B3 IIIe		& 40	& 193 & ... \\
NGC\,346-023\tablefootmark{*} & B0.2e	& 130 & 50  & ... & 		NGC\,330-070 & B0.5e 		& 39 & 348 & 1 \\
NGC\,346-024 & B2e:shell	& 160 & 190 & ... &             	NGC\,330-076 & B3e		& 35 & 62  & 1 \\
NGC\,346-036 & B0.5 Ve		& 142 & 287 & ... &            	NGC\,330-083 & B3 IIIe		& 38 & 140 & ... \\
NGC\,346-041 & B2e		& 130 & 144 & 1 &        	NGC\,330-085 & B3:e		& 38 & 191 & 1 \\
NGC\,346-045 & B0.5 Vne	& 135 & 181 & ... &            	NGC\,330-087 & Be-Fe 		& 24 & 214 & 1 \\ 
NGC\,346-048 & Be (B3 shell)  & 121 & 158 & 2 &          	NGC\,330-091 & B0e 		& 32 & 272 & 1 \\
NGC\,346-061 & B1-2e 		& 118 & 336 & 1 &         	NGC\,330-096 & B1-3e		& 26	& 175 & 1 \\
NGC\,346-064 & B1-2e		& 117 & 108 & 1 &           	NGC\,330-100 & Be (B0-3) 	& 25 & 373 & 1 \\
NGC\,346-065 & B3e		& 105 & 222 & 1 &           	NGC\,330-112 & B1-3e 	 	& 24 & 262 & 1 \\
NGC\,346-067 & B1-2e 		& 105 & 351 & 1 &		&&&&  \\
NGC\,346-068 & B0 Ve 		& 109 & 378 & 1 &           	NGC\,2004-023 & B2e 	  	& 169 & 102 & ... \\
NGC\,346-069 & B1-2e 	 	& 102 & 186 & ... &            	NGC\,2004-025 & B2e	  	& 153 & 83  & ... \\
NGC\,346-072 & B1-2e 	 	& 112 & 102  & ... &            	NGC\,2004-027 & B0e 		& 148 & 182 & ... \\
NGC\,346-073 & B1-2e 	 	& 110 & 190 & ... &           	NGC\,2004-035 & B1e 		& 125 & 244 & ... \\
NGC\,346-076 & B2e 		& 106 & 237 & 1 &          	NGC\,2004-039 & B1.5e	   	& 106 & 212 & ... \\
NGC\,346-089 & B1-2e	 	& 91 & 79  & ... &             	NGC\,2004-048\tablefootmark{*} & B2.5e	& 86 & 244 & ... \\ 
NGC\,346-091 & B1e 		& 77 & 49  & ... &              	NGC\,2004-056 & B1.5e	    	& 74 & 229 & ... \\ 
NGC\,346-095 & B1-2e 	 	& 97 & 227 & 1 &             	NGC\,2004-058 & O9.5 Ve	& 125 & 195 & ... \\
NGC\,346-096 & B1-2e 	 	& 100 & 343 & 1 &           	NGC\,2004-067 & B1.5e	    	& 56 & 237 & ... \\
NGC\,346-110 & B1-2e	 	& 93 & 243 & 1 &             	NGC\,2004-083 & B1.5:e     	& 73 & 194 & 1 \\
&&&&&                                                                              	     	NGC\,2004-089 & B2.5e	    	& 67 & 288 & ... \\
NGC\,330-025\tablefootmark{*} & B1.5e 	& 93 & 250 & ... &            	NGC\,2004-092 & B2e	    	& 61 & 171 & ... \\ 
NGC\,330-029 & B0.2 Ve		& 85 & 209 & ... &            	NGC\,2004-096 & B1.5e	    	& 63 & 245 & ... \\
NGC\,330-031 & B0.5 Ve		& 98 & 178 & ... &              	NGC\,2004-115\tablefootmark{*} & B2e	& 111 & 15  & 3 \\
NGC\,330-034 & B1-2e 		& 54 & 231  & 1 &            	&&&&  \\
NGC\,330-044 & B1-2e	 	& 25 & 184 & 1 &             	N\,11-073	& B0.5e			& 143 & 222 & ... \\
NGC\,330-050 & B3e 		& 67 & 214 & 1 &                N\,11-074	& B0.5e			& 149 & 102 & ... \\
NGC\,330-054 & B2e 		& 44 & 147 & 1 &		N\,11-078 	& B2e 		& 153 & 93   & 2 \\ 
NGC\,330-060 & B2.5e 		& 37 & 88  & 1 &              	N\,11-081 	& B0:ne 		& 135 & 363 & 1 \\ 
NGC\,330-062 & B3e  		& 43 & 241 & 1 &            	N\,11-090\tablefootmark{*} 	& B2e	        	& 119 & 118 & ... \\ 
NGC\,330-064 & B3:e 		& 23 & 269 & 1 &           	&&&&  \\ 
\hline
\end{tabular}
\tablefoot{Spectral types and projected rotational velocities (\vsini)  are taken from \citet{hun08a}, unless stated otherwise.
\tablefoottext{*}{Value of \vsini has been redetermined (see Sect. \ref{s_ew})}
\tablefoottext{1}{Omitted: No reliable measurements of the \ion{Mg}{ii}/\ion{Si}{iii} or \ion{He}{ii} spectrum}
\tablefoottext{2}{Omitted: The presence of a strong shell spectra}
\tablefoottext{3}{Omitted: Not a classical Be-type star. Short period binary.}
}
\end{center}

\end{table*}

\subsection{Equivalent width measurements} 	\label{s_ew}

Equivalent widths (EWs) for metal absorption lines in narrow lined spectra can normally be reliably measured by fitting Gaussian profiles to the observed features \citep[see, for example][]{tru07}.  However as noted in \citet{hun09a}, for stars with projected rotational velocities, \vsini $>$ 50\kms, the dominant broadening mechanism is rotation.  Hence it was necessary to utilise rotationally broadened profiles in order to reliably estimate the equivalent widths of our sample of Be-type stars.

The high rotational velocities of Be-type stars lead to additional complications when  analysing the strength of metal lines.  \citet{vra97} and \citet{por03} discuss these and in particular that the continuum can mask the outer wings of features present in the spectra of such objects.  This can affect the estimation of the continuum level, line width and the line strength, normally resulting in a systematic underestimation of these quantities.  To address this issue, \citet{hun08a} deduced projected rotational velocities from the helium spectrum which is less susceptible to such effects due to its intrinsic strength.  As noted in Table \ref{t_sample}, six objects have had their rotational velocity estimates re-determined and these have been used throughout the current analysis.  This was because rotationally broadened profiles generated using the estimates of \citet{hun08a} did not reliably reproduce the observed metal lines.  Typically the difference was $\sim$30\% of the original value.

\begin{figure}[hbtp]
\begin{center}
\includegraphics[angle = 90,scale=0.35]{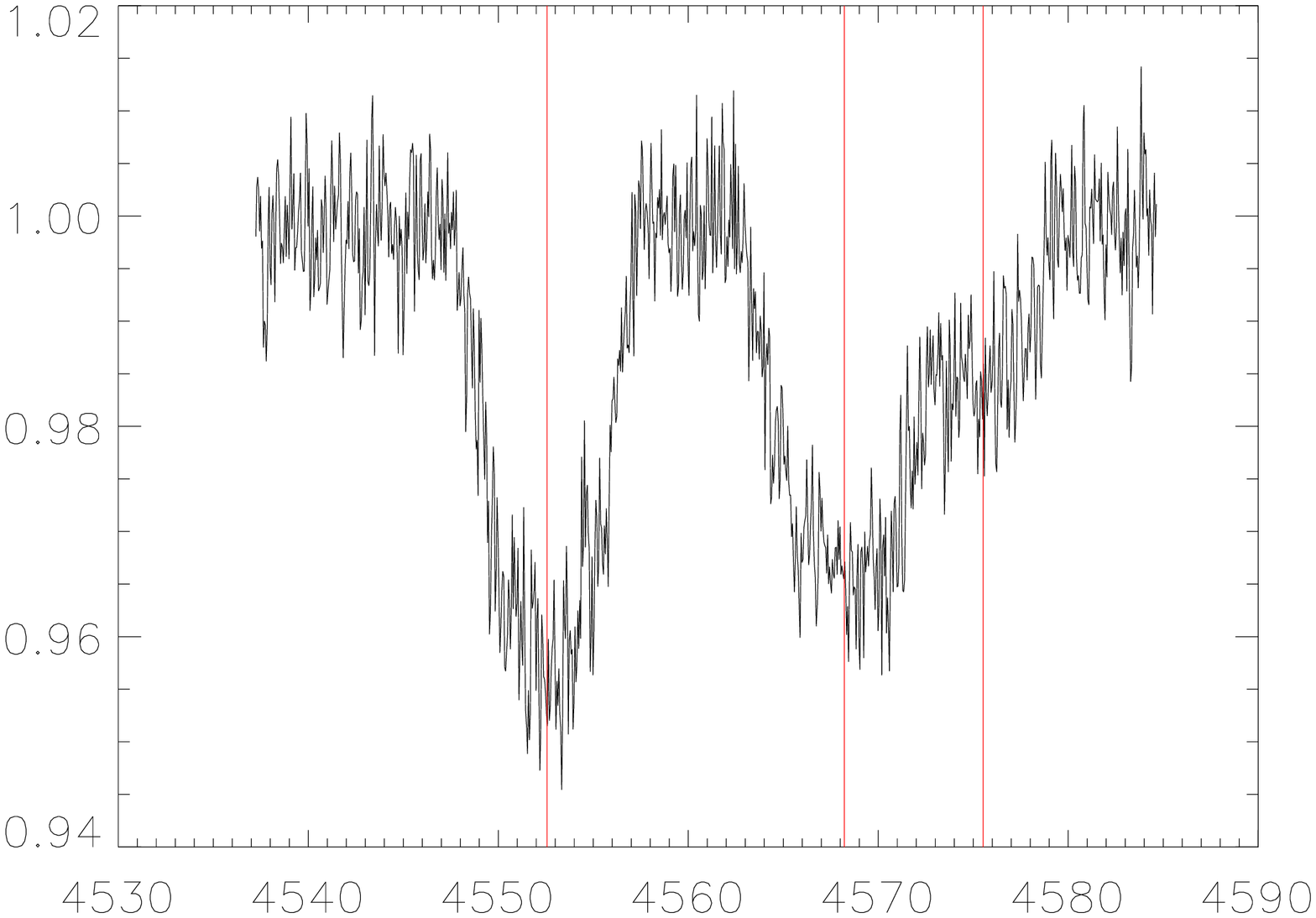}
\caption{Blending of \ion{Si}{iii} lines in NGC\,346-008, which has a \vsini $\sim$ 299\kms. Red lines indicate line centroids as chosen by the fitting procedure (for fixed relative wavelengths of the triplet).}
\label{f_SiIII}
\end{center}
\end{figure}

Equivalent width measurements of metal lines were then undertaken using the methodology discussed by \citet{hun07}.  Briefly, single or multiple rotational broadened profiles generated with the projected rotational velocities listed in Table \ref{t_sample} were fitted to the observed spectra using a least-squares technique. The significant broadening can cause blending with other lines, which can also affect the equivalent width measurements.  The effects of such blending is illustrated for the star, NGC\,346-008  (with \vsini$\sim 299$\kms) in Fig. \ref{f_SiIII} for the \ion{Si}{iii} triplet near 4560\AA.  Lines were excluded from the subsequent analysis if rotationally broadened profiles that had been constrained to the appropriate wavelength centroids did not yield a satisfactory fit.

Table \ref{t_lines} lists the equivalent width estimates for selected lines of nitrogen, magnesium and silicon.  NGC\,346-096 has been included in this table due to the measurement of \ion{N}{ii} line, but the lack of reliable \ion{Si}{iii}/\ion{Mg}{ii} spectra meant it was excluded from further analysis.  One of the primary aims of this paper is to determine reliable nitrogen abundances for our sample.  Hence for targets where  no \ion{N}{ii} line at 3995\AA\ could be distinguished, upper limits to this equivalent width were estimated.  The methodology adopted has been modified from the approach discussed in \citet{hun07}, as follows.  On a star by star basis, a rotationally broadened profile was added to the observed spectrum and its strength varied until it was visible within the spectral noise.  The line was then measured and the resulting equivalent width (rounded to the nearest 5 m\AA) was taken as a conservative upper limit.


\begin{table*}
\caption{Equivalent width measurements and associated uncertainty estimates for 32 Be-type stars.} \label{t_lines}
\centering
\begin{tabular}{lccccccc}
\hline\hline
Star & \ion{N}{ii} 3995 & \ion{Si}{iii} 4552  & \ion{Si}{iii} 4567  & \ion{Si}{iii} 4574  & \ion{Si}{iv} 4116  & \ion{Si}{ii} 4130  & \ion{Mg}{ii} 4481  \\
 & (m\AA) & (m\AA) & (m\AA) & (m\AA) & (m\AA) & (m\AA) & (m\AA) \\
\hline

NGC\,346-004 & 133$\pm$7 & 211$\pm$10 & 154$\pm$8 & 84$\pm$5 & 83$\pm$5   & ... & 61$\pm$4 \\
NGC\,346-008 & 65 $\pm$7 & 274$\pm$15   & 216$\pm$12   & 106$\pm$6   & ... & ... & 112$\pm$6 \\
NGC\,346-009 & 42 $\pm$4 & 101$\pm$8   & 81$\pm$6   & 49$\pm$4   & 163$\pm$13   & ...& 31$\pm$2 \\
NGC\,346-023 & 31$\pm$5 & 74$\pm$13   & 52$\pm$9   & 34$\pm$6   & 49$\pm$9   & 24$\pm$4    & ...  \\
NGC\,346-024 & 71$\pm$14 & 68$\pm$14   & ...  & 50$\pm$10   & ... & 87$\pm$18    & 135$\pm$28 \\
NGC\,346-036 & 68$\pm$14 & 88$\pm$18  & 117  $\pm$22  & 106$\pm$21   & ... & ... & ... \\
NGC\,346-045 & 60$\pm$8 & 93$\pm$11   & 77$\pm$9   & 52$\pm$6   & ... & 36$\pm$10   & 52$\pm$6 \\
NGC\,346-069 & $\leq$40 & 99$\pm$18   & 68$\pm$13   & 41$\pm$14   & ... & 63$\pm$12   & 52$\pm$14 \\
NGC\,346-072 & $\leq$35 & 48$\pm$10   & 27$\pm$8 & 16$\pm$9   & ... & ... & 51$\pm$10 \\
NGC\,346-073 & $\leq$40 & 71$\pm$13   & 63$\pm$14   & 36$\pm$11   & ... & ... & 57$\pm$8 \\
NGC\,346-089 & $\leq$25 & 70$\pm$7   & 53$\pm$13   & 46$\pm$5   & ... & ... & ... \\
NGC\,346-091 & $\leq$25 & 75$\pm$11   & 58$\pm$8   & 46$\pm$5   & ... & ... & 45$\pm$6 \\
NGC\,346-096 & 120$\pm$17 & ... & ... & ... & ... & ...& ... \\
\\
NGC\,330-025 & 56$\pm$10  & 171$\pm$10 & 105$\pm$11 & 70$\pm$10 & ... & ... & 101$\pm$13 \\
NGC\,330-029 & $\leq$50 & 53$\pm$14  & 30$\pm$11 & 27$\pm$10  & 38$\pm$8   & ... & ...   \\
NGC\,330-031 & $\leq$40 & 145$\pm$30  & 87$\pm$18 & 47$\pm$10  & 51$\pm$10  & ... & ...   \\
NGC\,330-069 & $\leq$90 & 73$\pm$13	& 46$\pm$13 & ... & ... & ... & 160$\pm$30 \\
\\
NGC\,2004-023 & 39$\pm$8     & 78$\pm$15     & 74$\pm$15     & 25$\pm$6     & ... & ... & 61$\pm$12 \\
NGC\,2004-025 & 24$\pm$5     & 74$\pm$15 & 63$\pm$13 & 32$\pm$6 & ... & ... &  99$\pm$19 \\
NGC\,2004-027 & 54$\pm$3     & 102$\pm$6    & 76$\pm$5     & 59$\pm$5     & ... & ... & ... \\
NGC\,2004-035 & 75$\pm$15     & 29$\pm$6     & 57$\pm$11     & 30$\pm$6     & ... & ... & 90$\pm$17 \\
NGC\,2004-039 & 52$\pm$6     & 133$\pm$16    & 110$\pm$13    & 49$\pm$7     & ... & ... &  195$\pm$23 \\
NGC\,2004-048 & 57$\pm$8     & 116$\pm$13  & 108$\pm$12    & 70$\pm$10    & ... & ... &  202$\pm$22 \\
NGC\,2004-056 & 53$\pm$14     & 129$\pm$30   & ...          & ...          & ... & ... & 119$\pm$17 \\
NGC\,2004-067 & 83$\pm$15     & 163$\pm$29   & 135$\pm$24   & 94$\pm$17    & ... & ... & 135$\pm$24 \\
NGC\,2004-089 & 36$\pm$10    & 110$\pm$14   & 94$\pm$12    & 60$\pm$11    & ... & ... & 182$\pm$24 \\
NGC\,2004-092 & $\leq$40	   & 104$\pm$14    & 96$\pm$12    & 69$\pm$22     & ... & ... & 114$\pm$30 \\
NGC\,2004-096 & 54$\pm$16     & 144$\pm$43   & 101$\pm$30   & 77$\pm$23    & ... & ... & 82$\pm$25 \\
\\
N\,11-073 & 32$\pm$9 & 54$\pm$15 & ... & ... & ... & ... & 52$\pm$14 \\
N\,11-074 & 30$\pm$6 & 84$\pm$16 & 92$\pm$18  & 81$\pm$16   & 58$\pm$12   & ... & ...  \\
N\,11-090 & 19$\pm$5 & 67$\pm$14 & 68$\pm$14  & 34$\pm$7   & ... & ... &  52$\pm$11 \\
\hline
\end{tabular}
\tablefoot{
The \ion{Si}{iii} and \ion{Mg}{ii} absorption lines listed were necessary in the methodology used to obtain atmospheric parameters (Sec. \ref{s_atmos}).  \ion{N}{ii} at 3995\AA\, was used to compare our sample with evolutionary models (Sec. \ref{s_evo}).
}
\end{table*}



\section{Analysis}                              \label{s_analysis}

\subsection{Non-LTE atmosphere calculations}                      \label{s_nlte}

The analysis in this paper follows that of previous work on the FLAMES dataset, making use of a B-type stellar non-LTE model atmosphere grid, computed using the {\sc tlusty}  model atmosphere codes and the line formation code {\sc synspec} \citep{hub88, hub95, hub98}. The grid used covers the effective temperature range 12\,000 -- 35\,000\,K in steps of 1\,500\,K at its lowest resolution and 500\,K at its highest resolution; surface gravity covers the range $4.5$\,dex down to the Eddington limit in steps of $0.25$\,dex.  The microturbulent velocities have discrete values of 0, 5, 10, 20, and 30\kms.  Further details can be found in \citet{rya03} and \citet{duf05}.

\subsection{Limitations of the Be-type stellar analysis}			\label{s_diff}

The nature of Be-type stars as discussed in the introduction are inconsistent with homogeneous, plane parallel, static model stellar atmospheres.  The presence of a circumstellar-like disc, short-term variations and high rotation rates of possibly up to $\sim90\%$ of the critical break-up velocity are characteristics that will affect the geometry and structure of the star but are not  accounted for within the {\sc tlusty} grid.  For this reason the approach taken here is relatively simplistic, but a differential analysis should be sufficient to constrain the large range of nitrogen enrichments in our sample and provide some insights even in the light of these known issues.

\subsection{Stellar atmospheric parameters}                      \label{s_atmos}

A static stellar atmosphere is  characterised by the four parameters: effective temperature (\teff), surface gravity (\logg), microturbulent velocity ($\xi$) and chemical composition.  Due to the interdependence of these properties an iterative procedure must be adopted when using the {\sc tlusty} grids in order to obtain estimates for each parameter.  For our targets, we used the appropriate SMC or LMC metallicity grid and initially assumed that the continuum contribution from a circumstellar disc was negligible.  We will subsequently refine our atmospheric parameters by attempting to allow for any such contribution.

\subsubsection{Effective temperature}            \label{s_teff}

Effective temperatures were estimated using three different methods depending on the spectral quality, degree of rotational broadening and temperature regime.   For objects with effective temperatures greater than 24\,000\,K, the \ion{He}{ii} absorption line at 4686\AA\, has a high temperature sensitivity and hence was adopted as our primary methodology.  Other \ion{He}{ii} lines at 4541\AA\, and 4199\AA\, could also have been used, but in many of our objects neither of these lines were well observed.  Therefore for consistency it was decided to only model the feature at  4686\AA.  The observed line profile was compared with theoretical spectra (corrected to allow for rotational broadening) from within the {\sc tlusty} grids and the temperature estimated using a $\chi^{2}$ technique.  The quality of the observed spectra typically leads to a typical uncertainty in \teff of $\pm 500\,K$ but there were several objects for which larger uncertainties were appropriate due to the relatively poor quality of the agreement between observation and theory.  As a note of caution, this method of determining \teff is dependent on the helium abundance of the object in question.  The \teff values quoted for the objects in Tables \ref{t_method} \& \ref{t_results} assume a normal helium abundance, but we can make an estimate to how the increase in helium abundance due to rotational mixing may affect the surface temperature.  Although He abundances have only been calibrated for slowly rotating evolutionary models\citep{heg00}, \citet{bro11a} suggest a 10\% enrichment of helium is possible for fast-rotating B-type stars.  By using the {\sc tlusty} grid of model atmospheres we find that an equivalent increase in the strength of the \ion{He}{ii} 4686\AA\,absorption line leads to a temperature increase of $<$500\,K, hence we do not expect this to be a significant source of error.

The silicon ionisation balance was used for cooler objects without an observable \ion{He}{ii} spectrum, but where absorption lines due to \ion{Si}{iii} (at 4552\AA, 4567\AA\, and 4574\AA) and either \ion{Si}{ii} (at 4128\AA\, and 4130\AA) or \ion{Si}{iv} (at 4116\AA) were observed.  This methodology requires that the abundance estimates derived from the two ionisation stages agree.  For objects whose spectra contained neither \ion{He}{ii} nor two silicon ionisation stages, effective temperatures were estimated by requiring that the relative silicon to magnesium abundance was normal.  The former was deduced from the \ion{Si}{iii} multiplet at 4560\AA\, and the latter from the \ion{Mg}{ii} doublet at 4481\AA.  The baseline abundances were taken from \citet{hun07} and are given in Table \ref{t_baseabund}.  Eleven objects in the LMC used only this method, together with seven in the SMC.  

The use of three methodologies leads to the possibility of there being systematic discrepancies. Therefore where possible, we have compared for any given star the effective temperature estimates obtained using different methods. We note that this comparison was undertaken after the correction discussed in Sect. \ref{s_corrections} had been applied but our principal results would remain unchanged if we had used our initial estimates. In the first instance, we adopted the other atmospheric parameters (\logg and $\xi$) and estimate of disc contamination from our primary method and then redetermined only the effective temperature using the alternative methods. In Table \ref{t_method}, the effective temperature estimates are compared and in general, the agreement is reasonable particularly considering the caveats discussed in Sect. \ref{s_diff}.  For six stars, the typical difference between the \ion{He}{ii} and silicon ionisation balance was found to be 1\,300\,K with a standard deviation of 1\,300\,K.  This is consistent with the \ion{He}{ii}, silicon ionisation equilibrium comparison performed by \citet{kil91}. The comparison of the silicon ionisation balance and Si/Mg method was only possible in two cases but the values are in reasonable agreement.  A comparison was inappropriate (and indeed normally not possible) between the \ion{He}{ii} and Si/Mg methods, as the latter loses its temperature sensitivity above $\sim$25\,000\,K.  However there was one object, NGC\,346-004, which could be analysed using all three methodologies yielding estimates that had a range of 1\,500 K. 

Following this  comparison, a complete model atmosphere analysis (\teff, \logg, $\xi$ and $\Delta$[Disc]) was re-performed for each star using the alternative methods for estimating the effective temperature.  The results are again summarised in Table \ref{t_method}. The difference in effective temperature estimates from the \ion{He}{ii} and Si equilibrium methods was now on average 1\,700\,K with a standard deviation of 1\,100\,K.    For the silicon ionisation equilibrium and Si/Mg methods  the difference in one case was larger than in the previous comparison. For the one star, where all three methods could be utilised, the range of effective temperature estimates was 1\,600 K.

\begin{table}
\centering
\caption{Present-day chemical composition of the Milky Way \citep{nie11},  LMC and SMC \citep[][]{hun07}.  Abundances are presented on the scale $12+\log[X/H]$.}
\label{t_baseabund}
\begin{tabular}{lccc}
\hline \hline
	& MW & LMC & SMC \\
\hline
C & 8.35 & 7.75 & 7.35 \\
N & 7.82 & 6.90 & 6.50 \\
O & 8.77 & 8.35 & 8.05 \\
Mg & 7.57 & 7.05 & 6.75 \\
Si & 7.50 & 7.20 & 6.80 \\
\hline
\end{tabular}
\end{table}

For all three methods, observational uncertainties imply a typical error of $\pm 500\,$K.  However there are likely to be significant additional uncertainties due to the limitations in the theoretical methods (see Sect. \ref{s_diff}), which are very difficult to quantify.  Where more than one method has been available, the estimates normally agree to within $\pm 2\,000\,$K implying that this might be a reasonable  estimate of the effective temperature uncertainty.


\begin{table*}[!]
\caption{Comparison of atmospheric parameters and element abundances using different methods to estimate the effective temperature. }
\label{t_method}
\centering
\begin{tabular}{lccccccccccc}
\hline \hline
Star          & \vsini & Method of & Method &  \teff  & \logg & $\xi$ & Si & Mg & N &  $\Delta$[Disc]\\
              &  (\kms) & Analysis & Constraints & (K) &  & (\kms) & & & & $\%$  \\
\hline
NGC\,346-004 & 266 & \ion{He}{ii} & Primary & 24000 & 2.90 & 10 & 6.8 & 6.8 & 7.8 & 0 \\
NGC\,346-004 & 266 & Si & Fixed & 23500 & ... & ... & 6.7 & 6.8 & 7.7 & ... \\
NGC\,346-004 & 266 & Si & Free &  23400 & 2.90 & 10 & 6.7 & 6.8 & 7.7 & 0  \\
NGC\,346-004 & 266 & Si/Mg & Fixed & 25000 & ... & ... & 6.9 & 6.9 & 8.0 & ... \\
NGC\,346-004 & 266 & Si/Mg & Free &  22400 & 2.80 & 11 & 6.7 & 6.7 & 7.6 & 0 \\
\\
NGC\,346-009 & 181 & \ion{He}{ii} & Primary & 29500 & 3.40 & 10\tablefootmark{*} & 6.9 & 6.5 & 7.6 & 0  \\
NGC\,346-009 & 181 & Si & Fixed & 28900 & ... & ... & 6.8 & 6.5 & 7.5 & ... \\
NGC\,346-009 & 181 & Si & Free & 28800 & 3.40 & 10\tablefootmark{*} & 6.8 & 6.5 & 7.5 & 0 \\
\\
NGC\,346-023 & 50   & \ion{He}{ii} & Primary & 31000 & 3.65 & 10\tablefootmark{*} & 6.8 & ... & 7.6 & 0\\
NGC\,346-023 & 50   & Si & Fixed & 27400 & ... & ... & 6.2 & ... & 6.9 & ... \\
NGC\,346-023 & 50   & Si & Free & 29200 & 4.00 & 5\tablefootmark{*} & 6.9 & ... & 7.3 & 35 \\
\\
NGC\,346-024 & 190 & Si & Primary & 16500 & 2.50 & 20\tablefootmark{*} & 6.7 & 6.5 & 7.3 & 0  \\
NGC\,346-024 & 190 & Si/Mg & Fixed & 16800 & ... & ... & 6.6 & 6.6 & 7.3 & ... \\
NGC\,346-024 & 190 & Si/Mg & Free & 21000 & 3.20 & 10\tablefootmark{*} & 6.7 & 6.7 & 7.5 & 40 \\
\\
NGC\,330-029 & 209 & \ion{He}{ii} & Primary & 31000 & 4.40 & 1 & 6.8 & ... & 7.8 & 25 \\
NGC\,330-029 & 209 & Si & Fixed & 30700 & ... & ... & 6.8 & ... & 7.7 & ... \\
NGC\,330-029 & 209 & Si & Free & 30700 & 4.40 & 1 & 6.8 & ... & 7.7 & 25\\
\\
NGC\,330-031 & 178 & \ion{He}{ii} & Primary & 28000 & 3.50 & 17 & 6.8 & ... & 7.3 & 35 \\
NGC\,330-031 & 178 & Si & Fixed & 25900 & ... & ... & 6.6 & ... & 7.1 & ... \\
NGC\,330-031 & 178 & Si & Free & 26100 & 3.65 & 23 & 6.7 & ... & 7.2 & 50 \\
\\
N\,11-074	& 102 & \ion{He}{ii} & Primary & 27500 & 3.65 & 10\tablefootmark{*} & 7.2 & ... & 7.2 & 40 \\
N\,11-074 & 102 & Si & Fixed & 26800 & ... & ... & 7.2 & ... & 7.2 & ... \\
N\,11-074	& 102 & Si & Free & 25100 & 3.40 & 10\tablefootmark{*} & 7.2 & ... & 7.1 & 50 \\
\hline
\hline
\end{tabular}
\tablefoot{
The method used is indicated as follows: \ion{He}{ii}: profile fitting of the \ion{He}{ii} absorption line at 4686\AA; Si: silicon ionization equilibrium, requiring at least 2 ionisation stages of Si; Si/Mg: uses the relative abundance estimates of silicon and magnesium see Sect. \ref{s_teff} for more details.  The estimated percentage contribution of disc emission ($\Delta$[Disc]) to the continuum is also listed.  The parameters from the primary methodology are listed first together with those from other effective temperature estimators, quoted as either `Fixed' or `Free'.  `Fixed' parameters are where other atmospheric parameters (\logg and $\xi$) and estimates of the disc contamination are taken from the primary method, whereas `Free' parameters allow all the parameters to change.
\tablefoottext{*}{Assumed values of microturbulence}
}
\end{table*}

\subsubsection{Surface gravity}          \label{s_logg}

Surface gravities were initially estimated by comparing the observed Balmer lines, H$_\mathrm{\gamma}$ and H$_\mathrm{\delta}$ with theoretical spectra from the {\sc tlusty} grids; the estimates derived from both lines normally agreed to within $0.2$\,dex.  Throughout our sample, the hydrogen line profiles tended to exhibit strong narrow absorption or emission features at the line centre.  \citet[][and references therein]{por03} discuss the structure of line profiles in Be-type stars in relation to the kinematical nature of the circumstellar disc.  The presence of a disc can result in single peaked or double peaked line profiles that may also be asymmetric and this was more prominent in the lower series members,  H$_\mathrm{\alpha}$\ -- H$_\mathrm{\gamma}$. Hence it was decided to adopt the estimates from the H$_\mathrm{\delta}$ transition.  The typical formal error in the fitting procedure was less than $\pm 0.2$\,dex but there will be additional uncertainties due to the simplifications in our methodology.

The use of  different methods to derive the effective temperature (see Table \ref{t_method}) led to a mean difference in the gravity estimates of 0.04\,dex with a standard deviation of 0.14\,dex for the six stars analysed by the \ion{He}{ii} and silicon ionisation equilibrium methods.  Hence this would not appear to be a major source of uncertainty and we adopt a typical uncertainty in these estimates of $\pm 0.2$\,dex.

\subsubsection{Microturbulence}    \label{s_xi}

The microturbulent velocities were derived from the \ion{Si}{iii} multiplet at 4552\AA, 4567\AA\,and 4574\AA, by ensuring that the abundance estimates did not depend on line strength.  For several objects,  it was not possible to obtain a value of microturbulence that fulfilled this criterium  and therefore an estimate of 0\kms has been adopted, which minimises this dependency.  This phenomenon  has been discussed in detail by \citet{hun07}, where it was found in the normal B-type stellar population in the Magellanic Clouds. For some stars it was not possible to deduce  a reliable microturbulence as, for example, only one to two lines in the \ion{Si}{iii} multiplet could be measured, or the lines lay near to the linear part of the curve of growth. In these cases we have adopted values that are consistent with those found by \citet{hun09a} for stars with similar gravities, viz.   targets having a gravity between 3.5 -- 4.5\,dex were assigned $\xi$=5\kms; 3.0 -- 3.5\,dex, $\xi = 10$\kms;  2.5 -- 3.0\,dex, $\xi = 20$\kms.  These cases are noted in Tables \ref{t_method} and \ref{t_results}. 

The microturbulent velocities deduced using effective temperatures  derived from the \ion{He}{ii} profiles and the silicon ionisation equilibrium (see Table \ref{t_method}) differ by 4.0$\pm$5.0\kms, indicating the sensitivity of this quantity to the other atmospheric parameters. It is difficult to assess the reliability of these microturbulence estimates  given the assumptions discussed in Sect. \ref{s_diff}. However an uncertainty of the order of $\pm$5\kms would appear to be appropriate.

\subsection{Corrections}	\label{s_corrections}
The initial estimates of the atmospheric parameters have been refined to allow for two effects, namely blending of the \ion{Mg}{ii} doublet with a nearby \ion{Al}{iii} doublet and continuum radiation from a circumstellar disc.

\subsubsection{\ion{Mg}{ii} line blending}  	\label{s_al}
The procedure to estimate effective temperatures from the relative abundances of silicon and magnesium  requires a reliable measurement of the corresponding equivalent widths.  The \ion{Mg}{ii}  doublet is blended with an unresolved \ion{Al}{iii} doublet at 4479.9\AA\ with this blending being more significant at higher effective temperatures. We have attempted to correct for this blending as follows. FLAMES spectra of 30 narrow lined B-type stars within our four clusters, covering the effective temperature range $\sim$ 12\,000 -- 33\,000\,K \citep{hun07, tru07} were selected.  For these objects, it was possible to independently measure the equivalent widths of \ion{Al}{iii} and \ion{Mg}{ii} features and then obtain a correction factor which was the ratio of the \ion{Al}{iii} equivalent width to the sum of the two equivalent widths.

\begin{figure}[hbtp]
\centering
\includegraphics[scale=0.375,angle = 90]{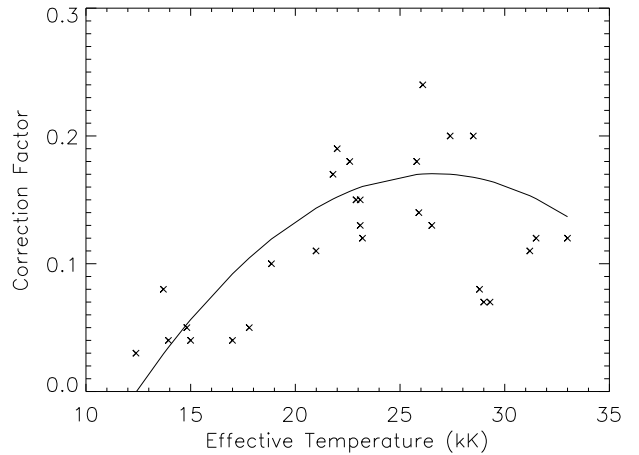}
\caption{Effective temperature is plotted against a correction factor, EW(\ion{Al}{iii})/EW(\ion{Al}{iii}+\ion{Mg}{ii}), derived from the \ion{Al}{iii} and \ion{Mg}{ii} absorption lines at 4479.9\AA\, and 4481.0\AA\,respectively.  This correction factor was obtained from the analysis for a sample of 30 B-type stars taken from \citet{hun07} and \citet{tru07} and based on the polynomial fit given by the solid black line.}
\label{f_CFvsTeff}
\end{figure}

In Fig. \ref{f_CFvsTeff}, this correction factor is plotted as a function of the effective temperature together with a polynomial least squares fit.   This was then used to scale the observed \ion{Mg}{ii} equivalent widths (see Table \ref{t_lines}) of our sample and hence deduce revised atmospheric parameter estimates for objects for which the Si/Mg method had been used. The effective temperature estimates were increased by typically $\sim$ 200\,K with a largest correction being $\sim$ 700\,K.  Hence this correction should not significantly increase the uncertainty in our effective temperature estimates.

\citet{lyu05} performed a detailed analysis of the \ion{Al}{iii}/\ion{Mg}{ii} blending in a comparable sample of slow rotating 
Galactic B-type dwarfs.  Comparison of our Fig. \ref{f_CFvsTeff} and Fig. 2 from of the work of \citet{lyu05} clearly illustrates that 
this blending effect is more prominent for the Milky Way sample than those in the LMC and SMC.  As both samples are of comparable size (28 for Lyubimkov and 
30 for this work) and span similar temperature ranges, it is reasonable to assume that the degree of blending is due to the 
metallicity environments.

\subsubsection{Correcting for disc contamination} 	\label{s_contamination}
The issue of addressing the spectral contribution from a circumstellar disc around a Be-type star is not straightforward. We have adopted a relatively simplistic approach that makes the following assumptions:
\begin{itemize}
\item{the contribution from the circumstellar disc is a featureless continuum, that is a constant fraction of the stellar continuum over the spectral range considered.}
\item{being main sequence objects, our targets will have the same silicon abundance as normal B-type stars in the same cluster.}
\end{itemize}
Silicon abundances were estimated from the \ion{Si}{iii} triplet for all our targets and were compared with the baseline abundances (Table \ref{t_baseabund}) for the SMC and LMC. The  difference ($\delta$Si) was assumed to be due to the disc continuum and an iterative procedure was developed to allow for this.  Initially, a fractional value for the disc continuum ($C$) was estimated from $\delta$Si\ and the equivalent widths of all metal lines were then scaled by the reciprocal of $(1-C)$.  These revised estimates of the stellar equivalent widths were used to estimate the effective temperature for stars using the silicon ionisation balance or Si/Mg methods.  For stars with temperatures estimated using the \ion{He}{ii} spectrum, the line profiles were corrected  by subtracting the estimated fractional contribution and renormalising the feature.  A similar procedure was applied to the H$_\mathrm{\delta}$ lines and the surface gravity re-estimated.  To complete the iteration, microturbulences were re-calculated using the corrected silicon line strengths.  With these new atmospheric parameters, silicon abundances were re-estimated, resulting in a revised disc contribution.  This iteration procedure was repeated until the values $\delta$Si were within $\pm0.1$\,dex. After applying these corrections to those targets where the disc contribution was non-zero, the typical increase in effective temperature was  $\sim 1\,600K$, the increase in surface gravity was $\sim 0.45$\,dex and the increase in microturbulence was  $<$\,5\kms.

\subsubsection{Photospheric abundances}	\label{s_photoabund}

Table \ref{t_results} summarises the photospheric abundance estimates (or upper limits) for our targets with all values rounded to the nearest 0.1\,dex.  Our principal aim has been to try and constrain the evolutionary status of our Be-type sample and hence their nitrogen abundances are crucial. The methodology has also led to estimates of the silicon and in some cases magnesium abundances. However these are not expected to be significantly affected by mixing from a hydrogen burning stellar core.  Additionally the spectra of these elements were used to deduce atmospheric parameters and the amount of continuum emission from a circumstellar disc and hence these abundance estimates cannot be considered to be independent.

\citet{fra10} used the \ion{N}{ii}  line at 3995\AA\ for a sample of supergiant stars and investigated how their abundance estimates differed from those deduced using all available \ion{N}{ii}  lines. They concluded that a conservative estimate of the {\it random error} was $\pm$0.15\,dex. Additionally for six stars with effective temperatures between 13\,000 and 25\,000 K, they found that the difference in the abundance estimate from the line at 3995\AA\ and the other \ion{N}{ii} lines was 0.03$\pm$0.08 dex. Hence we do not expect that the use of only one line should lead to significant systematic errors. As an additional test, the Be-type sample was searched for additional \ion{N}{ii} lines, with identifications being made in 11 stars. The other lines observed were 4630\AA, although this was often contaminated by an emission feature around 4629\AA, and 4447\AA; in two cases a line at 4601\AA\,was also visible. Including the abundance estimates from these features for objects with
\teff $<$26000 K  led to nitrogen abundances that were in general consistent with those estimated from the 3995\AA\,line alone, apart from
NGC\,346-004, where it was found to be 0.2 dex higher.  However as discussed in  Sect. \ref{s_nit}, this star exhibits shell like features thereby limiting the reliability of these measurements.

There could also be systematic errors due to, for example, errors in the atmospheric parameters. These have been extensively discussed by \citet{hun07} and vary depending on the strength of the feature and on the adopted atmospheric parameters. Typically such errors are of the order of 0.1 -- 0.2\,dex (but on occasions can be larger), which is consistent with the range of abundance estimates listed in Table \ref{t_method} for individual stars.

\begin{figure}
\centering
\includegraphics[scale=0.375,angle = 90]{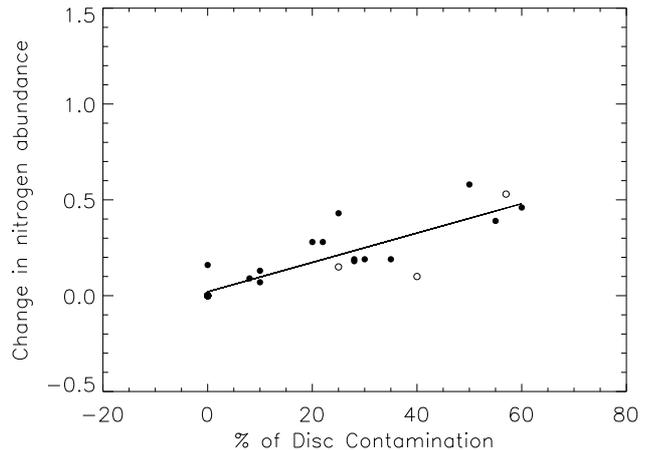} 
\caption{The change in nitrogen abundance after applying the contamination factor  is plotted against the estimated continuum contribution from the presence of a circumstellar disc . Both LMC and SMC stars are included (circles).  Stars showing shell-like features are noted as open circles.  A linear least squares fit (solid line) has been performed on all points and indicates a good correlation.}
\label{f_nitrogen_c}
\end{figure}

Combined these two sources of error would typically lead to an uncertainty of $\pm0.2$\,dex. However  there may be additional systematic errors due to for example the simplifications made in estimating the continuum contamination due to the disc. These are difficult to quantify but must be borne in mind when interpreting these results. For example, several of our targets appear to have negligible disc contamination despite their Be nature.   In Fig. \ref{f_nitrogen_c} we show the effect that this disk contamination has on our derived nitrogen abundances and one can see an approximate correlation such that every error of 10\%\ in the disc contamination results in an error of 0.1\,dex in nitrogen abundance.  Therefore for some objects our uncertainties may be as high as 0.4\,dex.

\section{Discussion of results}	  	\label{s_discussion}

\subsection{Atmospheric parameters}                     \label{s_dis_atmos}

\begin{figure}
\centering
\includegraphics[angle=90,scale=0.4]{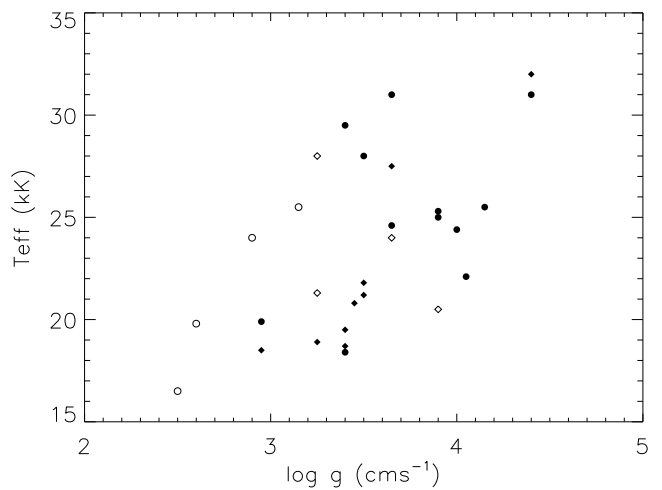}
\caption{Effective temperature plotted against surface gravity estimates for all stars listed in Table \ref{t_results}. Points are plotted for combined cluster samples; circles for the SMC and diamonds for the LMC.  Open symbols represent stars with shell-like features.}	
\label{f_teff_logg}
\end{figure}

Atmospheric parameters and photospheric abundances are presented for a total of 30 stars, 16 from the SMC and 14 from the LMC in Table \ref{t_results}.  The distribution of effective temperatures and gravities are shown in Fig. \ref{f_teff_logg} and indicate that the SMC and LMC targets span similar effective temperature ranges.  The surface gravities and microturbulent velocities also span similar ranges, although for the latter this may be due to the significant proportion (43\%) of adopted values (see Sect. \ref{s_xi}).  

\begin{table*}[!]
\centering
\caption{Estimates of the atmospheric parameters, disk contamination and silicon, magnesium, nitrogen abundances for the Be-type sample.}
\label{t_results}
\begin{tabular}{lcccccccrr}
\hline \hline
Star          & Method of & \vsini &   \teff  & \logg & $\xi$ & Si & Mg & N & $\Delta$[Disc]\\
                & Analysis    & (\kms) & (K) 	&  & (\kms) & & \\
\hline
NGC\,346-023 & \ion{He}{ii} & 50   & 31000 & 3.65 & 10\tablefootmark{\dagger} & 6.8 & ... & 7.6 & 0 \\
NGC\,346-009 & \ion{He}{ii} & 181 & 29500 & 3.40 & 10\tablefootmark{\dagger} & 6.9 & 6.5 & 7.6 & 0  \\
NGC\,346-089 & \ion{He}{ii} & 79   & 25500 & 4.15 & 0 & 6.9 & ... & $\leq$7.4 & 10\\
NGC\,346-036\tablefootmark{*} & \ion{He}{ii} & 287 & 25500 & 3.15 & 10\tablefootmark{\dagger} & 6.8 & ... & 7.0 & 0\\
NGC\,346-045 & Si/Mg & 181 & 25300 & 3.90 & 5\tablefootmark{\dagger} & 6.9 & 6.7 & 7.3 & 20\\
NGC\,346-091 & \ion{He}{ii} & 49   & 25000 & 3.90 & 0 & 6.8 & ... & $\leq$6.7 & 8\\
NGC\,346-073 & Si/Mg & 190 & 24600 & 3.65 & 0 & 6.8 & 6.8 & $\leq$7.1 & 10 \\
NGC\,346-069 & Si/Mg & 186 & 24400 & 4.00 & 8 & 6.8 & 6.7 & $\leq$7.0 & 30 \\
NGC\,346-004\tablefootmark{*} & \ion{He}{ii} & 266 & 24000 & 2.90 & 10 & 6.8 & 6.8 & 7.8 & 0  \\
NGC\,346-072 & Si/Mg & 102 & 22100 & 4.05 & 5\tablefootmark{\dagger} & 6.8 & 6.8 & $\leq$7.5 & 50 \\
NGC\,346-008\tablefootmark{*} & Si/Mg & 299 & 19800 & 2.60 & 20\tablefootmark{\dagger} & 6.8 & 6.7 & 6.9 & 0 \\
NGC\,346-024\tablefootmark{*} & Si & 190 & 16500 & 2.50 & 20\tablefootmark{\dagger} & 6.7 & 6.5 & 7.3 & 0 \\
&&&&&&&&& \\ 
NGC\,330-029 & \ion{He}{ii} & 209 & 31000 & 4.40 & 1 & 6.8 & ... & $\leq$7.8 & 25  \\
NGC\,330-031 & \ion{He}{ii} & 178 & 28000 & 3.50 & 17 & 6.8 & ... & $\leq$7.3 & 35 \\
NGC\,330-025 & Si/Mg  &250 & 19900 & 2.95 & 16 & 6.8 & 6.7 & 7.0 & 10 \\
NGC\,330-069 & Si/Mg  & 193 & 18400 & 3.40 & 10\tablefootmark{\dagger} & 6.8 & 6.8 & $\leq$7.8 & 0 \\
&&&&&&&&& \\ 
NGC\,2004-027\tablefootmark{*} & \ion{He}{ii} & 182 & 28000 & 3.25 & 10\tablefootmark{\dagger} & 7.2 & ... & 7.9 & 40\\
NGC\,2004-096\tablefootmark{*} & Si/Mg & 245 & 24000 & 3.65 & 7 & 7.1 & 7.0 & 7.2 & 25 \\
NGC\,2004-092 & Si/Mg & 171 & 21800 & 3.50 & 5\tablefootmark{\dagger} & 7.2 & 7.1 & $\leq$7.0 & 22  \\
NGC\,2004-067\tablefootmark{*} & Si/Mg & 237 & 21300 & 3.25 & 6 & 7.1 & 7.0 & 7.3 & 0 \\
NGC\,2004-023 & Si/Mg &102  & 21200 & 3.50 & 10\tablefootmark{\dagger} & 7.1 & 7.0 & 7.4 & 60  \\
NGC\,2004-056 & Si/Mg & 229 & 20800 & 3.45 & 10\tablefootmark{\dagger} & 7.1 & 7.0 & 7.2 & 28 \\
NGC\,2004-035\tablefootmark{*} & Si/Mg & 244 & 20500 & 3.90 & 16 & 7.2 & 7.0 & 8.0 & 57 \\
NGC\,2004-025 & Si/Mg & 83   & 19500 & 3.40 & 14 & 7.2 & 7.0 & 7.1 & 55 \\
NGC\,2004-048 & Si/Mg & 244 & 18900 & 3.25 & 10\tablefootmark{\dagger} & 7.2 & 7.0 & 7.2 & 0  \\
NGC\,2004-039 & Si/Mg & 212 & 18700 & 3.40 & 26 & 7.1 & 7.0 & 7.3 & 28 \\
NGC\,2004-089 & Si/Mg & 288 & 18500 & 2.95 & 6 & 7.2 & 7.1 & 7.1 & 0 \\
&&&&&&& \\
N\,11-073	  & \ion{He}{ii} & 222 & 32000 & 4.40 & 5\tablefootmark{\dagger} & 7.2 & ... & 7.8 & 50 \\
N\,11-074	  & \ion{He}{ii} &102  & 27500 & 3.65 & 10\tablefootmark{\dagger} & 7.2 & ... & 7.2 & 40 \\
N\,11-090  & Si/Mg         & 118 & 24100 & 4.40 & 11 & 7.2 & 7.0 & 7.4 & 60 \\
\hline
\hline
\end{tabular}
\tablefoot{
\tablefoottext{*}{Indicating those objects showing shell-like spectral features}
\tablefoottext{\dagger}{Unable to obtain $\xi$ from \ion{Si}{iii} lines; value adopted using estimated gravity}
}
\end{table*}

\subsection{Nitrogen abundances} \label{s_nit}

For the SMC sample,  nitrogen abundance estimates have been obtained for Be stars with relatively high projected rotational velocities (greater than 200\kms), in contrast to the results of the B star sample of \citet{hun09a}, which normally had only upper limits.  Inspection of Table \ref{t_results} also reveals that these Be stars tend to have surface gravities lower than one would expect for (near) main sequence stars.
 
Therefore, we have reviewed the spectra of our Be-type sample and find the presence of previously over-looked shell-like Balmer lines in eight stars; NGC\,346-004, 008, 024, 036 and NGC\,2004-027, 035, 067, 096. Since we are presumably observing these stars {\em through} their equatorial discs it is likely that our simple assumption that we observing the sum of the photospheric spectrum and disc continuum is no longer valid, leading to an underestimate of the stellar gravities. Although it would be possible to correct these surface gravities,  the impact of absorption by the disc on the nitrogen spectrum is largely unknown.  Therefore we have not sought to correct the nitrogen abundances for these shell-stars but flag them in Fig.s \ref{f_nitrogen_c} \& \ref{f_teff_logg} as they may be unreliable.  A further eight stars may also show signs of disc absorption
(NGC\,346-009, 023, NGC\,330-025, 031,  NGC\,2004-023, 056, N\,11-073, 074) and their results should be regarded with some caution.

An apparent correlation between effective temperature and surface gravity is seen in Fig. \ref{f_teff_logg}, partially due to 8 objects with \logg = 2.5 -- 3.4 and with \teff$<$20\,000\,K, namely NGC\,346-008, NGC\,346-024, NGC\,330-025, NGC\,330-069, NGC\,2004-025, NGC\,2004-048, NGC\,2004-039, NGC\,2004-089.  The stellar parameters of these objects are consistent with objects towards the end (or beyond) their main sequence lifetime and might therefore show a stronger enrichment of Nitrogen.  As discussed in Sect. \ref{s_photoabund} there is a possibility that use of only the single \ion{N}{ii} 3995\AA\,line could result in an 
underestimation of the true nitrogen abundance.  Therefore these objects were again checked for additional N II lines.  Out of the 8 
objects, 4 showed at least one extra identifiable N II line, and by taking the averaged abundance from all measured N II lines
3 objects showed an enhancement to their nitrogen abundance of approximately 0.1 dex.  The remaining object is noted as having shell like features and hence the estimates may be unreliable.

The nitrogen abundance estimates in both metallicity regimes span significant ranges: SMC: 6.7 -- 7.8\,dex,  LMC: 7.0 -- 8.0\,dex.   The largest enrichments in both metallicity regimes are of a similar magnitude, being  $\sim$1.3 and $\sim$1.1\,dex, for the SMC and LMC respectively. For the SMC eight objects showed no evidence of the \ion{N}{ii} line at 3995\AA\, and the upper limits to their absolute nitrogen enhancements  (0.2 -- 1.2\,dex) indicate that at least in some cases the nitrogen enrichment must be small.  This is consistent with an analysis by \citet{len05} who found effectively no nitrogen enrichment for a pole-on Be-type star in the SMC cluster NGC\,330. Only one object (NGC\,2004-092) in the LMC yielded a nitrogen upper limit, which implied effectively no enrichment ($\leq$0.1\,dex).

\begin{figure*}[t!]
\centering
\includegraphics[scale=0.375,angle=270]{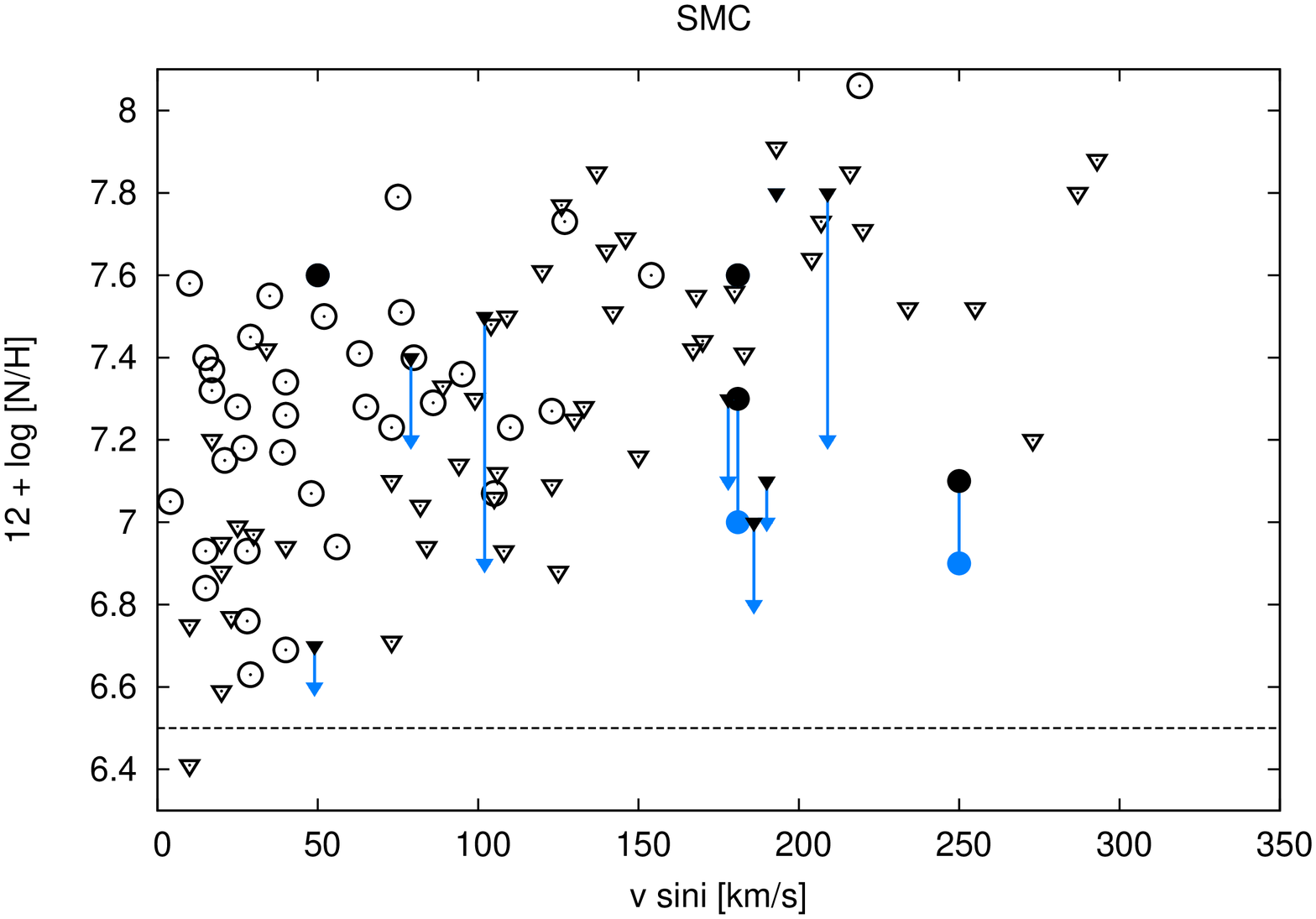} 
\includegraphics[scale=0.375,angle=270]{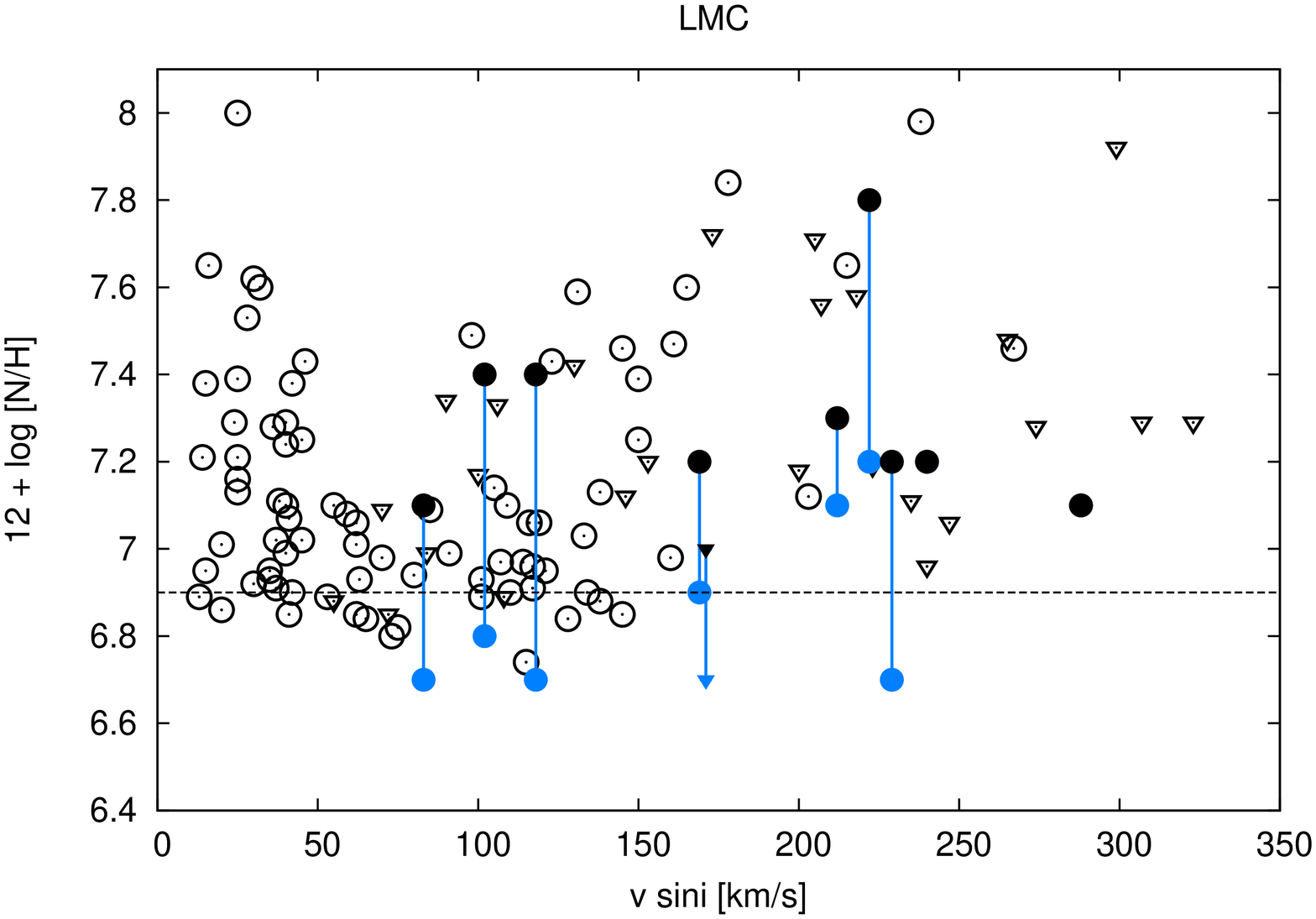} 
\caption{Comparison of Be-type stellar (black/blue) vsini and nitrogen abundances with those of B-type stars (black open circles) from \citet{hun09a} for both metallicity regions analysed (SMC - upper panel, LMC - lower panel).  Be-type star abundances are plotted for pre- (blue) and post-contamination (black) correction values.  Upper limits are denoted with triangles.}
\label{f_B_Be}
\end{figure*}

A surprising feature of our results is that the nitrogen abundances of normal B and our Be-type stars are found to have similar values.  The normal B-type stars are defined as those which exclude Group 1, N-normal fast rotators, and Group 2, N-rich slow rotating stars, \citep[as discussed by][]{hun09a}.  In Fig. \ref{f_B_Be}, the range of B-type and Be-type nitrogen abundances are given for the SMC and LMC.  For the SMC, there are a significant number of upper limits applied to the nitrogen measurements for fast rotating stars in both samples.  Nevertheless, this figure illustrates that even for the SMC, Be-type and B-type stars occupy a similar range of nitrogen abundances, and hint at a stronger enhancement at lower metallicity.  This comparison between Be- and B-type stellar abundances raises an important question; If Be-type stars are all fast rotators \citep[70\% of critical velocity,][and references therein]{por03} then why are they not more N-rich than is observed? We return to this in Sect. \ref{s_evo}.

As discussed previously one would consider more evolved giant and supergiant objects to exhibit a stronger enhancement to their 
nitrogen abundance.  \citet{lyu11} observed such nitrogen enhancements in A- and F-type supergiants which could have 
obtained their nitrogen enrichment from rotational mixing during a B-type MS lifetime.  To examine this further it would have been useful to investigate
the relationship between observed nitrogen enrichment and fundamental parameters such as the mass and main sequence lifetime of our objects.  
For Be-type stars, however, obtaining the intrinsic luminosity is not a straight forward task, due to the presence of circumstellar material 
and effects such as gravity darkening \citep{tow04}, and as a result we were unable to obtain any useful constraints. 

If we consider the relative nitrogen enrichments in reference to the cluster ages, NGC\,346 \& the N\,11 region are of roughly the same age of a few million years,
whilst  NGC\,2004 \& NGC\,330 are older but of comparable ages. However  we find there to be no correlation between the abundances estimates and {\em cluster} ages. As discussed by \citet{eva06}, the majority of the targets in the FLAMES survey are probably field stars, whose ages may well be different from that of the cluster.

\subsection{Projected rotational velocities} \label{v_eq}

\citet{hun08a} have previously discussed the projected rotational velocities for their FLAMES sample, including the  Be sample considered here. In particular they compared their results with  a previous study of \citet{mar06, mar07} of a large sample of B- and Be-type stars in NGC\,2004, and NGC\,330.

\begin{figure*}[t!]
\centering$
\begin{array}{cc}
\includegraphics[scale=0.375,angle=90]{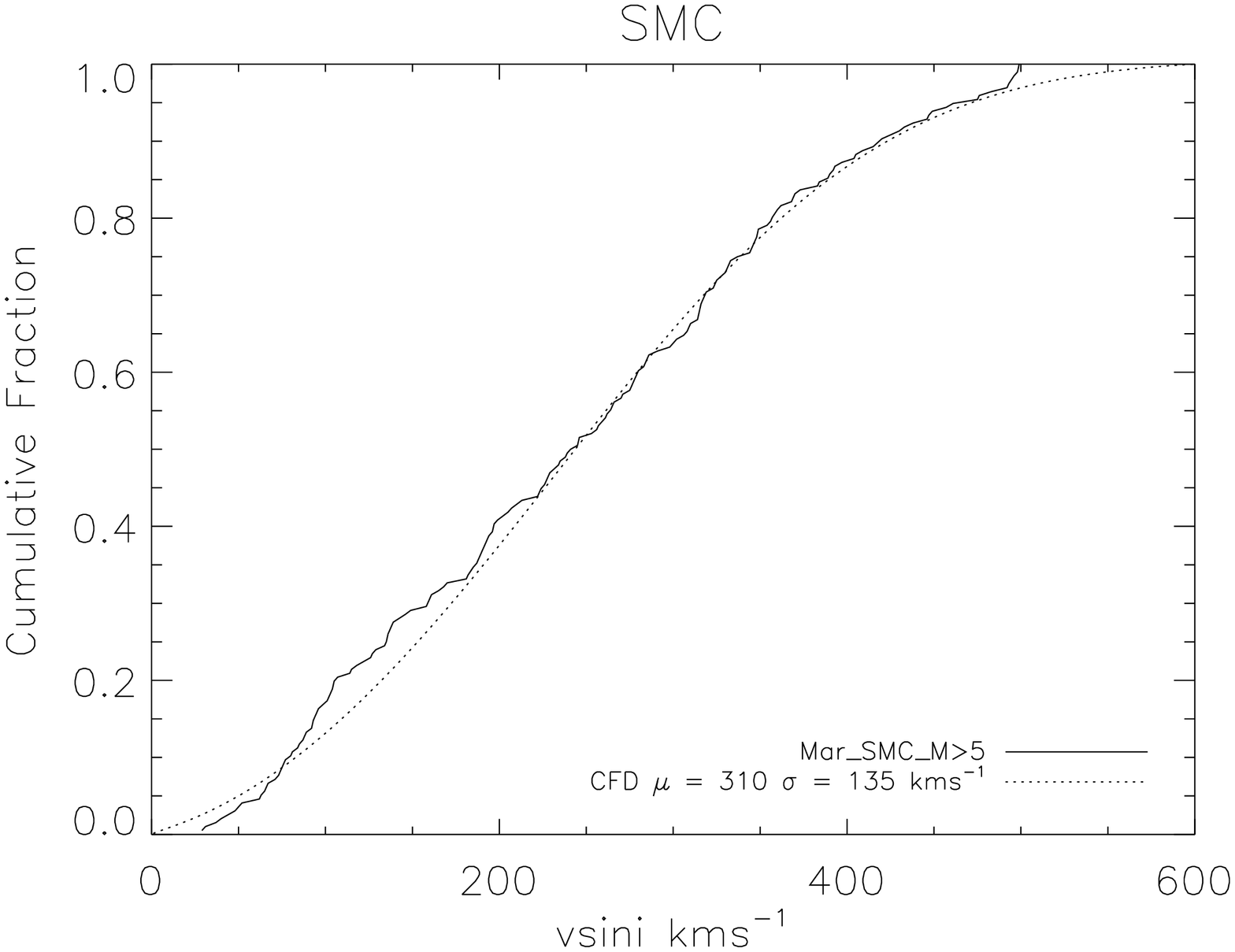} &
\includegraphics[scale=0.375,angle=90]{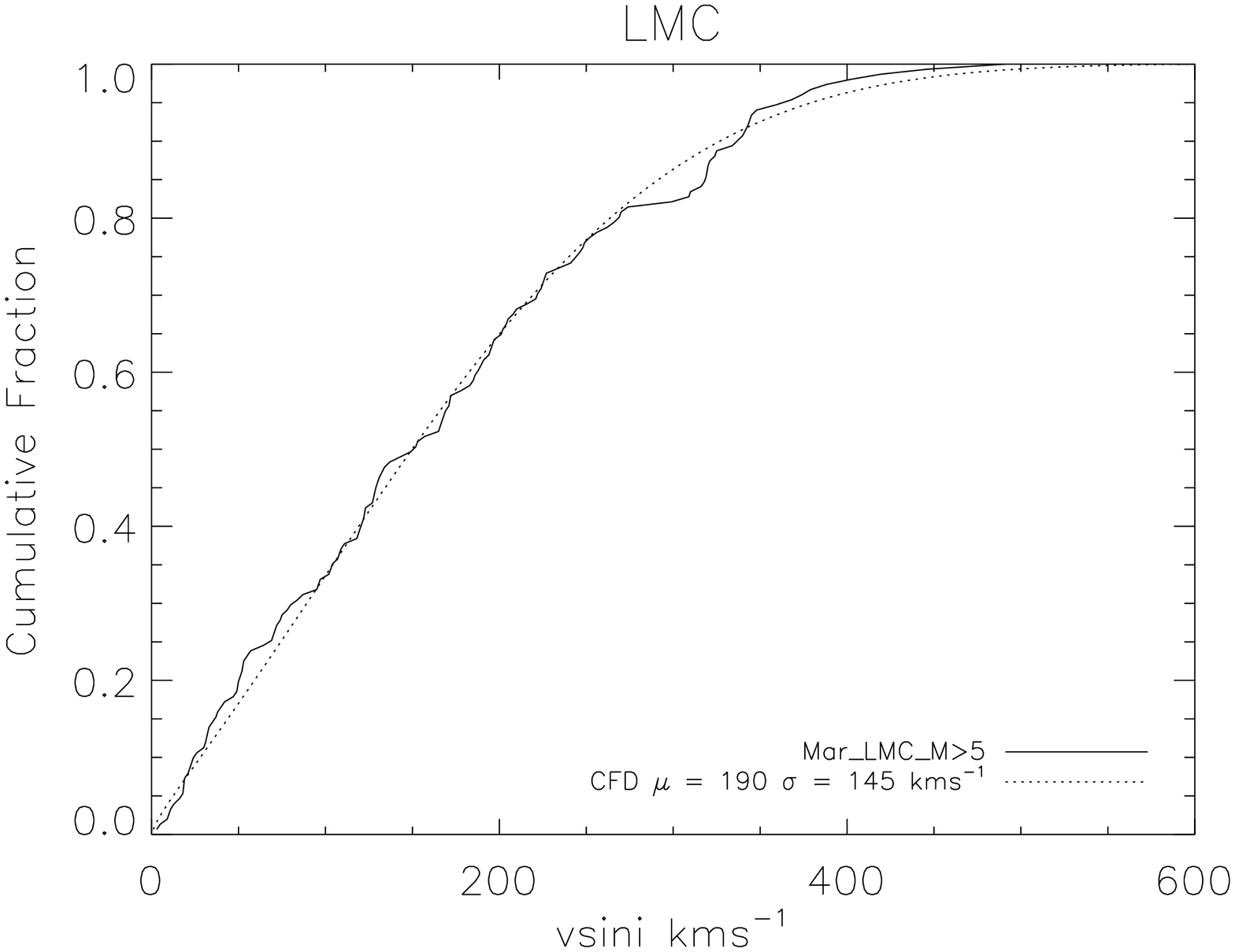} \\
\end{array}$
\caption{SMC ({\bf left}) and LMC ({\bf right}) samples from \citet{mar06, mar07} (solid line) fitted with Gaussian distributions (dashed line) in order to obtain estimates for the central velocity and width of the intrinsic rotational velocity.  Only objects greater than 5\Msun have been included in order to be comparable with our mass range.  The values adopted for the fits are given in Table \ref{t_both_sample}.}
\label{f_Mar_dist}
\end{figure*}

We have modelled our projected rotational velocities and those of \citeauthor{mar06} to infer the intrinsic rotational velocity distributions, which were assumed to follow a Gaussian profile - see \citet[see][]{hun08a} for details of the methodology.  Fig. \ref{f_Mar_dist} illustrates the results for the sample of \citeauthor{mar06}; the quality of the fit for our sample was somewhat worse, due to our smaller sample sizes.  The inferred rotational velocity distributions are summarised in Table \ref{t_both_sample} and imply that the majority of these Be-type stars are currently rotating at velocities that are considerably lower than their critical velocities.  From evolutionary models of  \citet[][]{bro11a} the critical velocity is typically 600 -- 650\kms for our mass range, which would suggest our objects have velocities $<$50\% of critical. We return to these results in Sect. \ref{s_evo}.

\begin{table}
\centering
\caption{Mean rotational velocities, $\frac{1}{e}$ widths and standard deviations  ($\sigma$) for an assumed Gaussian distribution  of rotational velocities.}
\begin{tabular}{lcccc}
\hline \hline
Sample	& n &  $v_\mathrm{0}$  & $\frac{1}{e}$ & $\sigma$ \\
		&	& (\kms)&  (\kms) &  (\kms) \\
\hline
This work: LMC & 19 & 210   & 100 & 70  \\
This work: SMC & 46 & 260   & 140 & 100  \\
\hline
Mar07: NGC\,2004     & 47	& 190 & 205 & 145 \\
Mar06: NGC\,330	& 131 & 310 & 190 & 135 \\
\hline\hline
\label{t_both_sample}
\end{tabular}
\tablefoot{
Velocities are considered for both the LMC and SMC by this work \citep[see][for further details]{hun09b} and that of \citeauthor{mar07}.  Given the uncertainties associated with these values they have all been rounded to the nearest 5\kms. Mar06 and Mar07 reference \citet{mar06, mar07} respectively.
}
\end{table}

\section{Comparison with evolutionary models}	\label{s_evo}

While the evolutionary status of Be-type stars is still uncertain it is interesting to compare our derived nitrogen abundances with the predictions of stellar evolution models
in the context of rotational mixing.  If one assumes that all Be-type stars are fast rotators, but are otherwise normal B-type stars, then clearly they should be important
tests of the efficiency of rotational mixing.  Here we compare our abundance estimates with those predicted from a simulated population of B-type stars based on theoretical evolutionary models of \citet{bro11a, bro11b}.  

\subsection{Model assumptions}

The code, STARMAKER, \citep{bro11b} simulates populations of early-type single stars, based on the evolutionary models presented in \citet{bro11a}, which take into account effects of rotation and mass loss.  STARMAKER predicts physical quantities (e.g. atmospheric parameters and abundances) for a prescribed sample of typically 1 -- 2 million of stars with initial mass, initial rotational velocity and age drawn from the corresponding distribution function (Salpeter IMF, initial distribution of rotational velocities, star formation history (SFH) and random inclination angle).  We adopted the parameters used in the simulations of \citet{bro11b} as these should be  appropriate to the current VLT-FLAMES sample.  In particular the underlying distribution of initial stellar rotational velocities reproduced the observed distribution of projected rotational velocities of the VLT-FLAMES B-stars.  As discussed in Sect. \ref{v_eq}, \citet{mar06, mar07} also observed a sample of B- and Be-type stars in NGC\,2004, and NGC\,330.  We note that the use of this distribution with our simulations do not change our results.

\begin{figure}[b!]
\centering
\includegraphics[angle=90,scale=0.4]{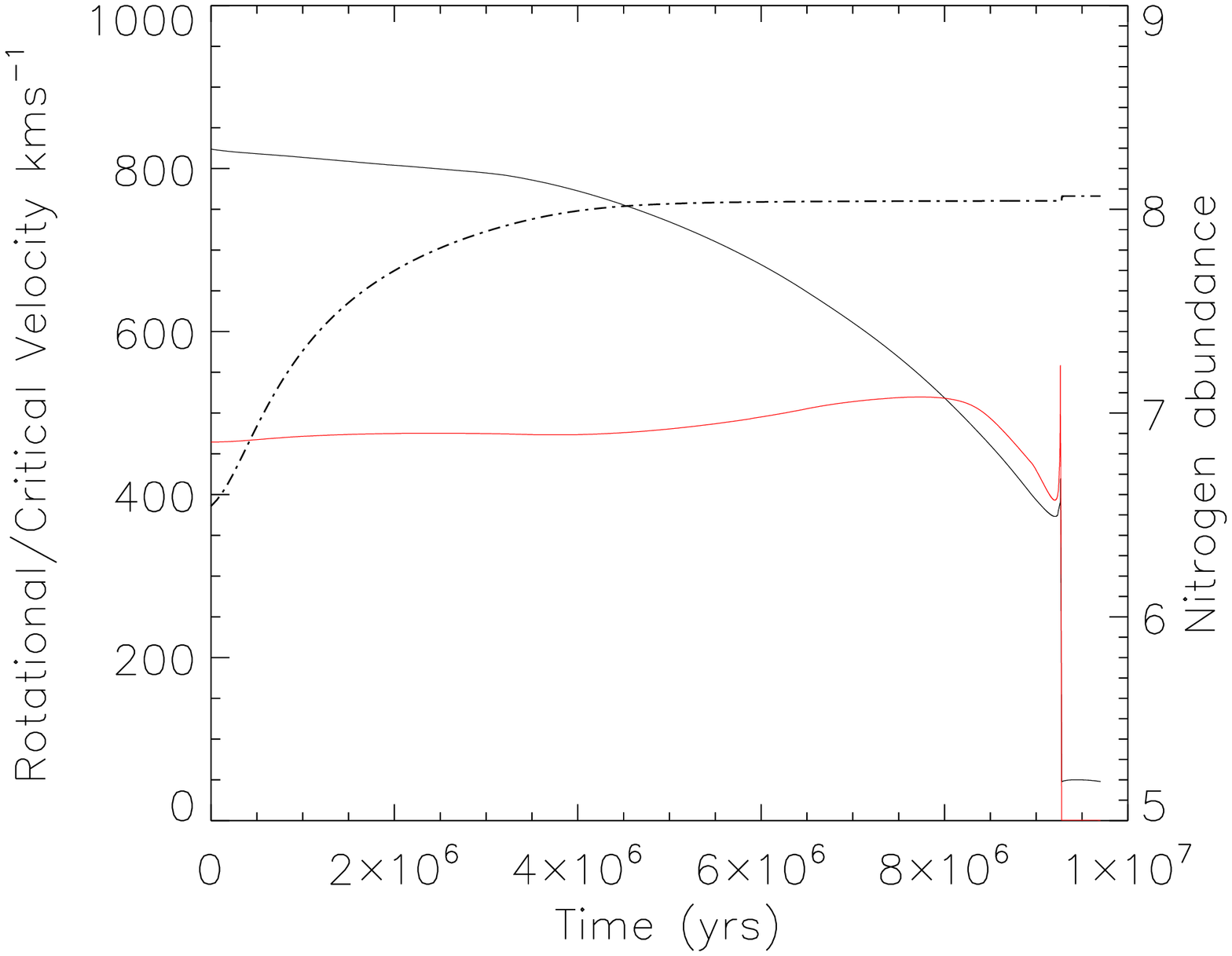}
\caption{Evolution of  a 20\Msun SMC B-type star with an initial rotational velocity $\sim$ 400\kms. The variation with time on the main sequence is shown for rotational velocity (red), critical velocity (black) and nitrogen enrichment (dashed black line) }
\label{f_track_20_400}
\end{figure}

The use of observed current projected rotational velocities to simulate the initial distribution of rotational velocities assumes that there is a relatively small change during the main sequence lifetimes. This is confirmed by the  simulations of \citet{bro11b} and illustrated in Fig. \ref{f_track_20_400}, which shows effectively no change in its rotational velocity during the main sequence lifetime for a twenty solar mass star. 

\subsection{Predictions}

Selection effects were applied to the simulated population in order to represent the observed sample. Typically $\sim$75\% of the simulated sample were removed to match the observed faint magnitude cut-off (i.e. exclusion of lower mass-stars), in addition to stars with ages beyond their main sequence lifetime and those with \teff$>$35\,000K and \logg $<$3.2\,dex. Other possible selection effects have been discussed in \citet{eva06} and \citet{bro11b}. However given the relatively simple nature of our comparisons we do not believe that these will be important.

\begin{figure}[t!]
\begin{center}
\includegraphics[bb = 80 50 540 770, angle = 270,scale=0.352]{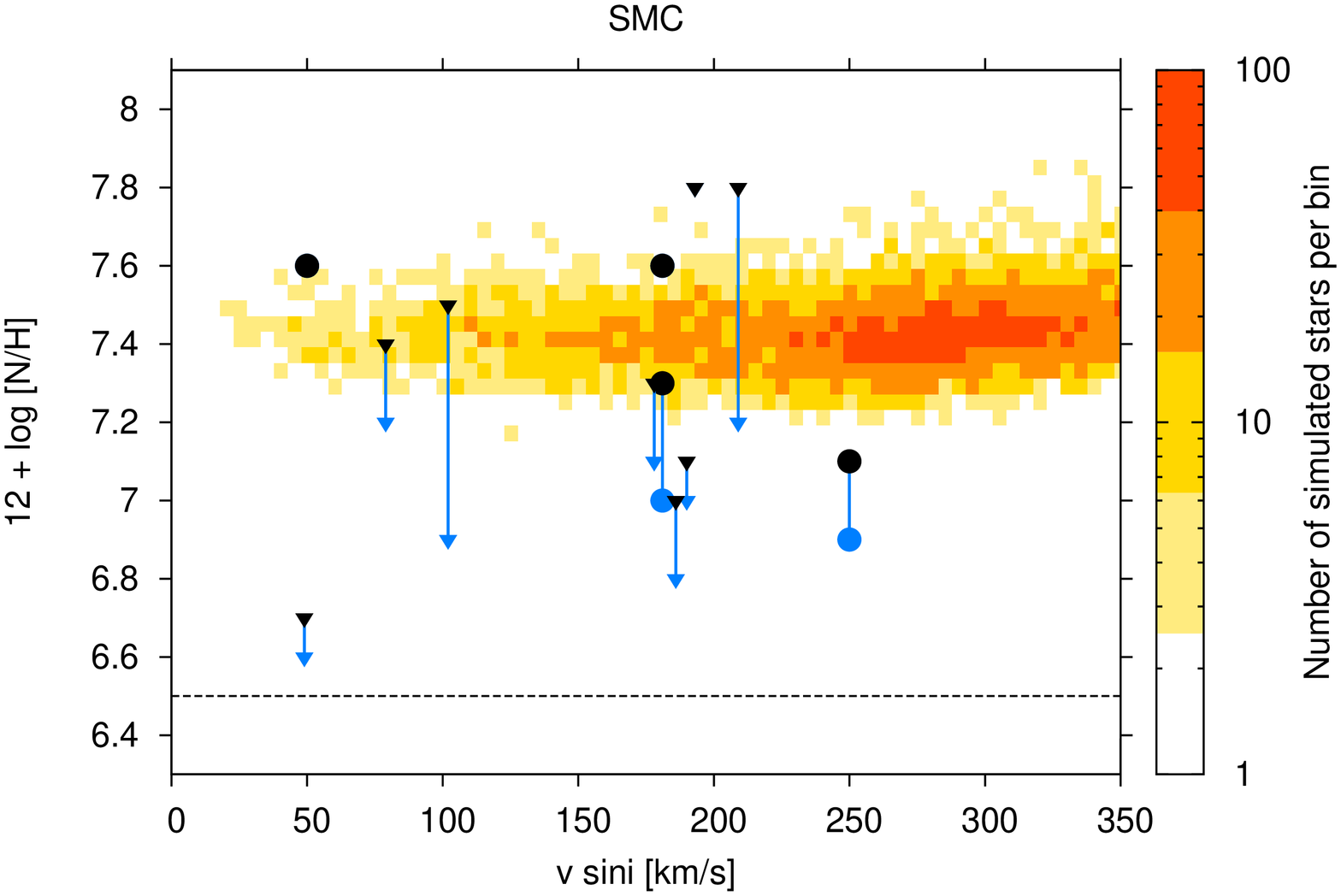}
\includegraphics[bb = 50 50 510 770, angle = 270,scale=0.352]{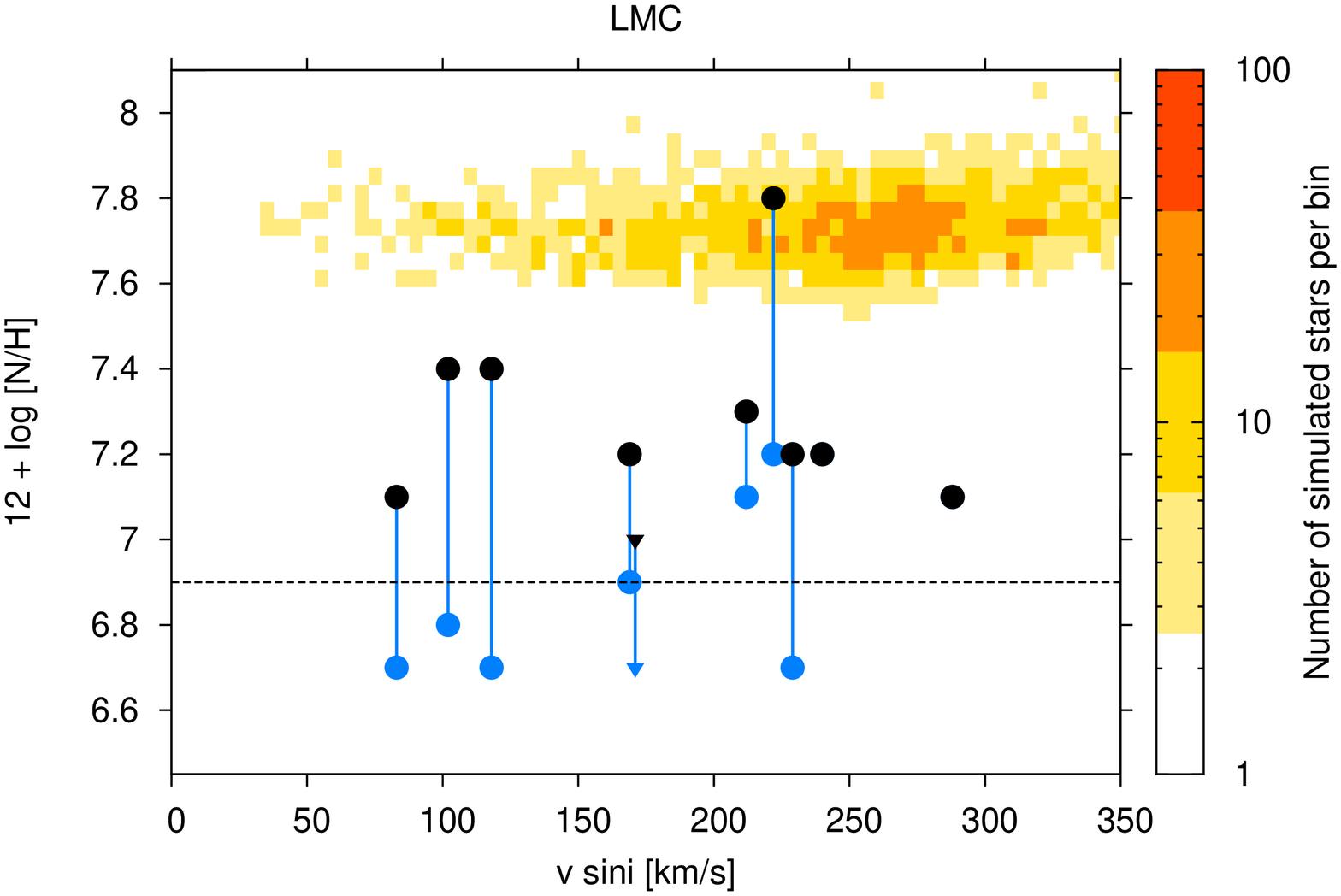}
\end{center}
\caption{Estimated nitrogen abundances and projected rotational velocities plotted for the SMC (top panel) and LMC (bottom panel) Be-type stellar samples before and after contamination correction (blue and black respectively).  Simulations are based on the rotational velocity distribution of Hunter et al.\ with only  B-type stars rotating with critical velocity between 65\% and 75\% as to represent a sample of stars rotating around 70\% of critical velocity.  Marked on each plot as a black dashed line is the baseline nitrogen abundance.  Simulations predict too high nitrogen abundances for the LMC, with a mixed view given for the SMC.}
\label{f_ev_70}
\end{figure}

\begin{figure}[hbtp]
\begin{center}
\includegraphics[bb = 80 50 540 770, angle = 270 ,scale=0.352]{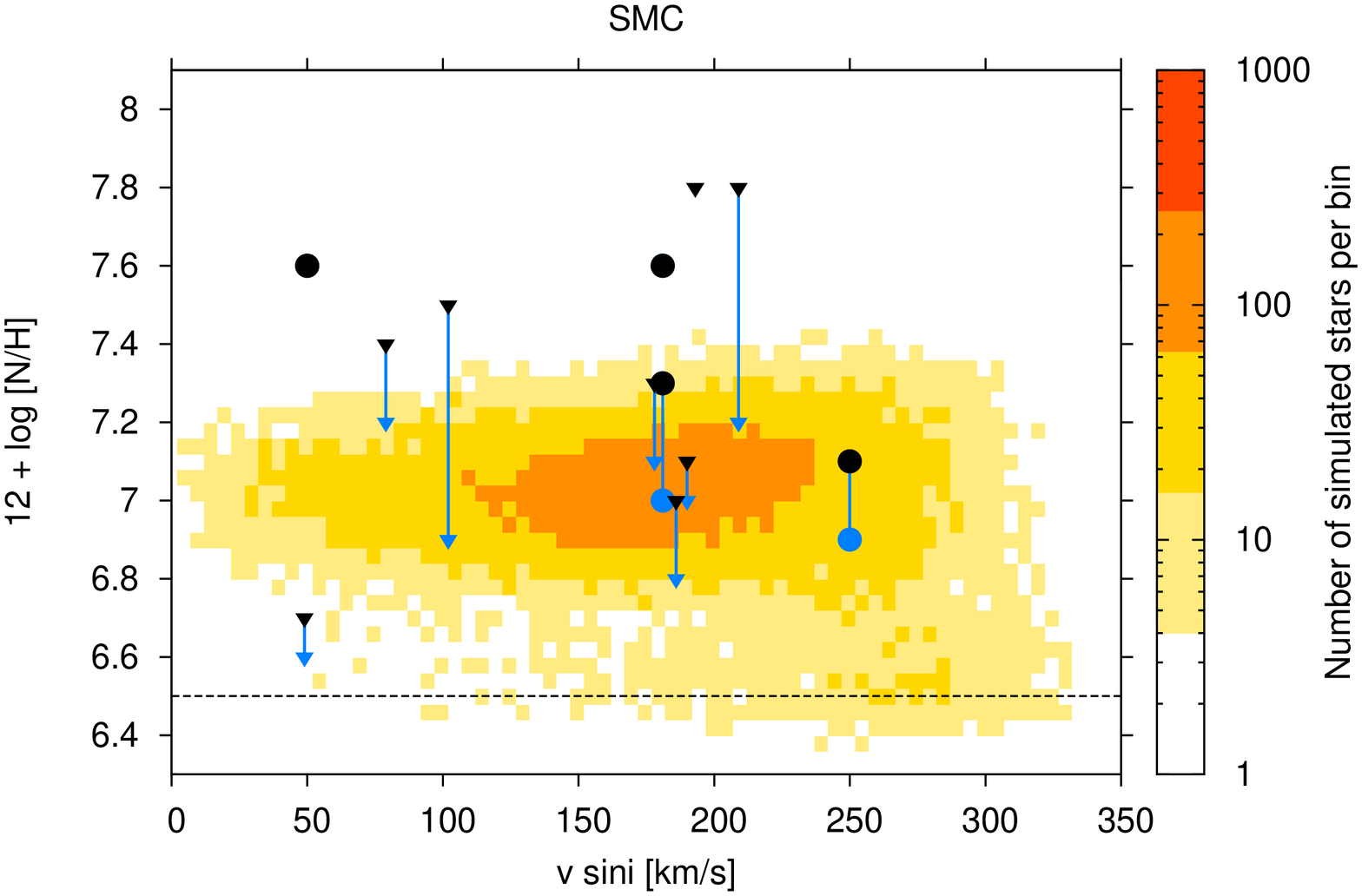}
\includegraphics[bb = 50 50 510 770, angle = 270 ,scale=0.352]{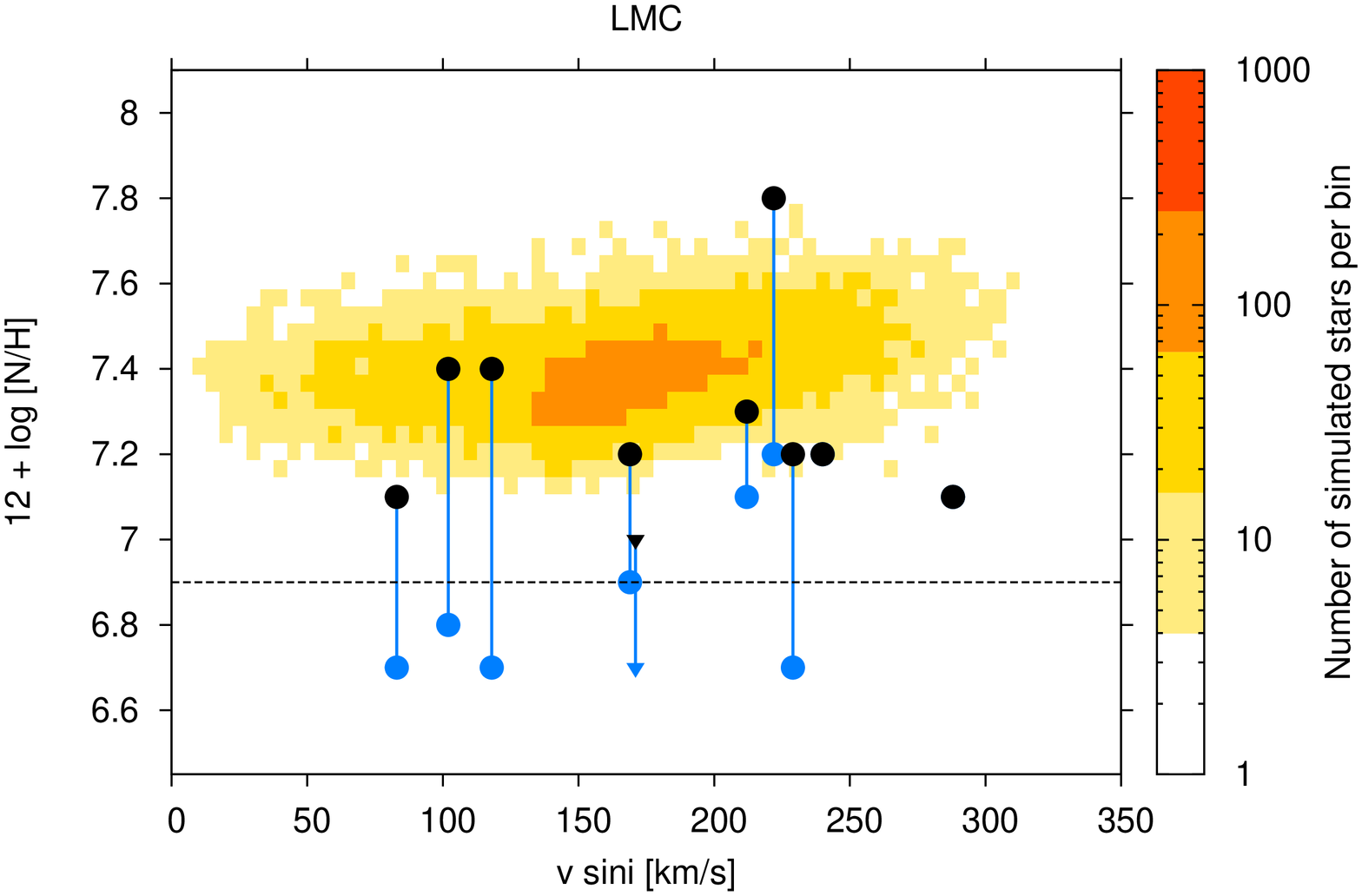} 
\end{center}
\caption{Estimated nitrogen abundances and projected rotational velocities plotted for the SMC (top panel) and LMC (bottom panel) Be-type stellar samples before and after contamination correction (blue and black respectively).  Simulations are based on the rotational velocity distribution of Hunter et al.\ with only  B-type stars rotating with critical velocity between 35\% and 45\% as to represent a sample of stars rotating around 40\% of critical velocity.  Marked on each plot as a black dashed line is the baseline nitrogen abundance.  Simulations predict a better agreement with LMC abundances and again a mixed view given for the SMC.}
\label{f_ev_40}
\end{figure}

With the assumption the population synthesis models should be able to reproduce our observed objects, we first investigate whether Be-type stars are a population of rapidly rotating B-type stars. In Fig.s \ref{f_ev_70} \& \ref{f_ev_40} simulations are presented assuming that Be-type stars rotate around 70\% or 40\% of their critical rotational velocity.  The simulations with 70\% clearly predict too high nitrogen abundances for the LMC, while for the SMC the picture is more mixed with some stars being well matched to the range of predicted abundances, while others are too low.  When we consider the simulations for stars at around 40\% of their critical velocity the LMC gives a better agreement, but again the SMC picture remains mixed and as already noted is complicated by the presence of more upper limits in the SMC.

Based on the nitrogen abundances, and assuming our evolutionary models are realistic, all Be-type stars do not rotate close to critical velocity as is often assumed.  This is consistent with the mean velocities found for different Be-type samples in Table \ref{t_both_sample}. Note that as discussed by \citet{mok06}, observed projected rotational velocities do not well constrain details of the distribution of the underlying rotational velocities. However they do constrain the mean rotational velocity and hence the values in Table \ref{t_both_sample} should be reliable.

Although there are significant uncertainties in our estimated nitrogen abundances as discussed in Sect. \ref{s_photoabund},  we believe that they are not compatible with stars that have spent most of their evolutionary history rotating at close to their critical velocity. As discussed above, simulations for such cases imply that effectively all the sample should have large nitrogen enhancements, which is not observed.  We can postulate different scenarios to explain our estimated nitrogen abundances, including: 
\begin{itemize}
\item On average Be-type stars rotate faster than normal B-type stars but have velocities significantly less than critical. 
\item If Be-type stars have had rotational velocities close to critical values, this phase must have been for a relatively short period of their main sequence lifetime.
\end{itemize}

For the first hypothesis, we are faced with the fundamental difficulty of why there are two types of stars. Certainly {\it on average} Be-type stars may have higher rotational velocities than B-type stars. However the range of velocities for Be-type stars implied by the $\sigma$-values in Table \ref{t_both_sample}  and for B-type stars in \citet{hun09a} imply that there will be a large overlap in the range of rotational velocities of the two groups. In turn this will imply that a substantial fraction of Be-type stars  have current rotational velocities that are less than normal B-type star mean velocity.  

If we assume that Be-type stars have experienced near critical rotation, we can use our grid of evolutionary models to infer an upper limit for the fraction of the main sequence lifetime. In Fig. \ref{f_track_20_400}, the evolution with time is shown for a 20\Msun SMC B-type star with an initial rotational velocity $\sim$ 400\kms. The nitrogen enrichment occurs relatively early in the main sequence lifetime, the majority in the first two million years (corresponding to less than one quarter of the main sequence lifetime). At this stage the nitrogen enhancement is more than 1.0\,dex corresponding to the largest enhancements found in our sample. Therefore if we take the average nitrogen enhancements found in our sample and we believe the theory of rotational mixing to be correct the star could only exist close to critical velocity for within the first ten percent of the main sequence lifetime. Consideration of other models within our grid or those of other investigations \citep{mey02} imply similar results with the nitrogen enrichment (and hence a short period of close to critical velocity) occurring early in the main sequence phase.  Alternatively, it is possible that the stars have undergone a recent spin up event, possibly due to an interacting binary system and that large-scale nitrogen enrichment (by rotational mixing) has not yet taken place.  This would require inefficient mass transfer as otherwise this could contribute to the gaining star's nitrogen abundance.

Recent work by \citet{hua10} discusses the rotation of Galactic B-type stars and infer a threshold for Be-type star phenomena by considering the maximum value of the rotational velocity to the critical velocity for their normal B-type star population. This is found to vary with stellar mass from $\sim$ 95\% for stellar masses  $<$ 4\Msun to $\sim$ 60\% for the higher masses ($>$ 8.6\Msun) appropriate to our sample.  Although we cannot explicitly determine at what rate of critical velocity our sample appears to be rotating at, from our population synthesis models and measured parameters the fraction of v$_\mathrm{crit}$ found by \citeauthor{hua10} would be consistent with Be stars in our mass range."


\section{Conclusions}                     \label{s_conclusions}

We have presented projected rotational velocities, atmospheric parameters and photospheric abundance estimates of three elements, including nitrogen for a small sample of Be-type stars in the SMC and LMC in order to compare with current evolutionary models. Our principal conclusions are:
\begin{itemize}
\item The estimated nitrogen enrichments are found to be similar for both Be- and B-type stars.  
\item Nitrogen abundances and projected rotational velocities have been compared with population synthesis models produced by \citet{bro11b}. They are inconsistent with a population that have spent most of their lifetime at near critical rotation velocities.
\item Be-type stars could have spent their main sequence lifetime rotating at velocities {\it on average} larger than B stars but still significantly less than the critical velocity. However the spread in velocities observed then implies that a substantial proportion of Be-type stars are currently rotating at velocities smaller than B stars.
\item Given the nitrogen enrichments found in our samples, Be stars could only experience near critical rotational velocities for a relatively short period of their main sequence lifetime (less than 10\%).
\end{itemize}

\begin{acknowledgements}
We are grateful to staff from the European Southern Observatory for assistance in obtaining the data.  We acknowledge financial support from the UK Science and Technology Facilities Council and the Department of Education and Learning in Northern Ireland.
\end{acknowledgements}



\end{document}